\documentclass[twocolumn]{aastex6}
\usepackage{amsmath,amssymb,amsfonts}
\usepackage{graphicx}
\usepackage{lipsum}
\usepackage{natbib}
\usepackage{color,soul}
\usepackage{hyperref}

\newcommand{\field}[1]{\textit{\textbf{#1}}}
\newcommand{\bay}{\texttt{BAYESTAR}}
\newcommand{\sqd}{sq.~deg.}
\newcommand{\spgr}{\textit{SpGr}}
\newcommand{\grar}{\textit{GrAr}}
\newcommand{\sear}{\textit{SeAr}}
\newcommand{\opar}{\textit{OpAr}}
\newcommand{\grin}{\textit{GrIn}}
\newcommand{\sein}{\textit{SeIn}}
\newcommand{\opin}{\textit{OpIn}}
\newcommand{\IUCAA}{Inter-University Centre for Astronomy and 
  Astrophysics, Post Bag 4, Ganeshkhind, Pune 411 007, India}
\newcommand{\WSU}{Department of Physics \& Astronomy, Washington State University,
1245 Webster, Pullman, WA 99164-2814, U.S.A}
\newcommand{\gransasso}{Gran Sasso Science Institute, Viale Francesco Crispi, 7, L'Aquila 67100, Italy}

\begin{document}

\title{An optimal method for scheduling observations of large sky error regions for finding optical counterparts to transients}

\author{Javed Rana\altaffilmark{1}, Akshat Singhal\altaffilmark{2}, Bhooshan Gadre, Varun Bhalerao, Sukanta Bose\altaffilmark{3}}
\affil{\IUCAA}
\altaffiltext{1}{javed@iucaa.in}
\altaffiltext{2}{\gransasso}
\altaffiltext{3}{\WSU}

\begin{abstract}


The discovery and subsequent study of optical counterparts to transient sources is crucial for their complete astrophysical understanding. Various gamma ray burst (GRB) detectors, and more notably the ground--based gravitational wave detectors, typically have large uncertainties in the sky positions of detected sources. Searching these large sky regions spanning hundreds of square degrees is a formidable challenge for most ground--based optical telescopes, which can usually image less than tens of square degrees of the sky in a single night. We present algorithms for optimal scheduling of such follow--up observations in order to maximize the probability of imaging the optical counterpart, based on the all--sky probability distribution of the source position. We incorporate realistic observing constraints like the diurnal cycle, telescope pointing limitations, available observing time, and the rising/setting of the target at the observatory location. We use simulations to demonstrate that our proposed algorithms outperform the default greedy observing schedule used by many observatories. Our algorithms are applicable for follow--up of other transient sources with large positional uncertainties, like {\em Fermi}--detected GRBs, and can easily be adapted for scheduling radio or space--based X--ray followup.
\end{abstract}




\section{Introduction}\label{sec:intro}

The coalescences of compact binary systems involving neutron stars or stellar mass black holes are among the most promising sources of gravitational waves (GWs) that the earth-based, broad-band interferometric GW detectors of the Advanced Detector Era (ADE) will detect. If at least one of the compact objects is a neutron star, the coalescence is also expected to give rise to electromagnetic radiation \citep[see for example][]{lp98}. In particular, electromagnetic (EM) signatures of mergers of binary neutron stars are conjectured to be the highly beamed short Gamma Ray Bursts \citep[SGRBs;][and references therein]{Eichler:1989ve,Berger2014}, and 
broadband EM transients \citep{mmd+10,Nakar2011, Metzger_Counterpart_2011}.
The rate of compact binary coalescences (CBCs) involving at least one neutron star detectable in the ADE of GW astronomy is projected to be 40-60 per year~\citep{aaa+10}. However, since only a small fraction of these sources will have an optimal orientation that maximizes its GW brightness on earth, an orientation that also favors the detection of a coincident on-axis afterglow, the expected rates of GW and EM coincident observations is as small as one in a few years~\citep{Metzger_Counterpart_2011,Petrillo2013a}. 


The scientific returns of joint EM-GW observations of CBCs, however, are expected to be highly significant. They will enrich our understanding of their EM beaming, energetics, and galactic environment, in addition to confirming if indeed CBCs involving neutron stars are the progenitors of SGRBs~\citep{bhh+09,Phinney2009}.
Moreover, the detection of a kilonova~\citep{tlf+13} concomitant with a GW event can shed light on the details of $r$-process nucleosynthesis in the unbound ejecta from a neutron star merger that powers the EM emission~\citep{Eichler_rProcess_2015, Metzger_nuclear_2014}.
These prospects make coincident EM-GW observations a high risk, but high gain pursuit. This is why all efforts must be made to enhance the probability of finding EM counterparts for every CBC event whose parameters do not disallow the possibility of it involving a neutron star.

Using only information from the interferometers, a GW source can be localized to a few hundred square degrees in the sky~\citep{Singer_Localization_2014}. Various groups from around the world then use high energy, optical, and radio telescopes to search for a coincident EM counterpart in this localization patch~\citep{aaa+14}. The primary challenge for counterpart searches is of scale: most optical telescopes have small fields of view ($\sim$ 0.1--1~\sqd)
\footnote{This may have biased many such telescopes from participating in EM followup for LIGO observing run O1, where many of the telescopes had wider fields of view~\citep{aaa+16}.}
, and take several minute long exposures, reaching $\sim$100 exposures per night. Thus, only a small fraction of the sky localization patch of a GW trigger can be covered by a given telescope in a night. Since the odds are stacked against finding an EM counterpart, it is important to optimize this follow--up imaging to maximize the probability of its discovery.

In the past a few different ways of gridding the large GW error regions (or ``patches'')
with tiles, each the size of a telescopes field of view, were explored. 
A few algorithms for ranking the various tiles in an error region, e.g., based on the probability of finding the source as deduced from GW parameter estimation methods, have been studied as well. The most prominent of these has been the ``greedy'' algorithm, which ranks the tiles purely on the basis of this probability, without accounting for any other factors, e.g., the setting time of the tiles. 
Higher ranked tiles are observed before lower ranked ones in order to maximize the probability of observing an EM counterpart of a GW event. However, past efforts discussing strategies for followup of such counterparts have some shortcomings. 
\citet{sps12} demonstrate that coordinated observations in a multi--telescope network increase the probability of imaging the EM counterpart location by upto a factor of two. They also find that accounting for visibility of the localization patch from a site reduces the overall probability of finding the true counterpart by roughly a factor of three. However, they do not discuss any way to reorder the observing sequence for setting, except perhaps implicitly in their simulated annealing method. Recently, \citet{chm+15} have addressed the trade--off between number of exposures and exposure time, but they too select a greedy algorithm where the highest probability tiles are to be imaged first. \citet{gsg+15} discuss some alternative methods of tiling and optimization for follow--up with telescopes with large fields of view.
%

In this paper, we describe multiple different algorithms for ranking tiles that remove some of the limitations of past work. We account for ground--based visibility constraints and schedule observations of tiles only when they are above the horizon. We explore the advantages of deviating from the default greedy algorithm. We also compare approaches for using a pre--defined grid of telescope pointings on the sky, as opposed to independent placement of each image. 
By applying these algorithms to thousands of simulations of realistic GW sky error regions or patches, and tile dimensions, we do a systematic comparison of these algorithms based on their probability of finding an EM counterpart.

The layout of the paper is as follows. In \S\ref{sec:all_algo} we describe seven algorithms for ranking and scanning tiles in sky patches. In \S\ref{sec:simulations} we perform extensive simulations using realistic GW sky patches
to compare the probabilities of finding the EM counterpart using these algorithms. We make specific recommendations on what observing strategies will be most optimal based on details of the sizes of tiles and the GW localization patches. 
In \S\ref{sec:discussion} we end with a discussion of application of our algorithms to other coincidence studies, the limitations of our algorithms and future scope for improvement.

\section{An optimal observing sequence for optical follow--up}\label{sec:all_algo}

In the early years of advanced gravitational wave detectors, the sky localization for any GW trigger will be rather coarse, spanning hundreds of square degrees on the sky \citep{Nissanke_Localization_2012,Singer_Localization_2014}. EM observers have to select a sky area for follow--up using an all--sky source location probability density function (PDF) calculated from the GW signal.
In this work, we select an area bounded by the contour that covers 95\% probability of containing the true location of the counterpart. Henceforth, we will refer to this area as the ``localization patch'', or simply ``patch''. Any follow--up requires taking images of parts of this patch using a telescope. We focus our attention on telescopes with a rectangular field of view, and call each such image a ``tile''. We do not discuss processing of these images for finding transients.

Imaging an area spanning several hundred square degrees is a formidable challenge for most observatories. Typical 1--2~meter class telescopes have fields of view that are a fraction of a square degree (cf. IUCAA Girawali Observatory: $13^\prime \times 13^\prime$), with few notable exceptions like the Palomar Transient Telescope (PTF, \cite{Law2009a}, $\sim$8 sq deg). The typical exposure times are no shorter than a few minutes --- limiting observers to around a hundred exposures a night, thereby making it impossible to image the entire patch. Observatories will thus need to select which parts of the patch they wish to follow--up, subject to various constraints like the diurnal cycle, telescope pointing limitations, available observing time, and the rising/setting of the target at the observatory location. We now propose algorithms to select and sequence tiles subject to these constraints, for maximizing the chances of an observatory finding the GW counterpart. 

\subsection{Defining the problem}\label{subsec:nom}

\newcommand{\texp}{\ensuremath{T_{\rm exp}}}

The aim of algorithms discussed here is to maximize the probability of finding the electromagnetic counterpart to the gravitational wave event. We assume that a single observation is sensitive enough to detect any EM counterpart that may be present within the field of view imaged by the telescope. Hence, we assume that repeat observations do not increase the probability of finding the EM counterpart. In practice, surveys often image each point multiple times --- this will be addressed in future work. The algorithms presented here are designed for a single telescope with a fixed imaging footprint on the sky. We do not discuss transient detection techniques.

Before we dive into the details of our algorithms, we list some terms that will be used in the following sections:
\begin{itemize}
  \item \textit{Patch}: Area of the sky bounded by a contour of constant probability density, such that it contains 95\% probability of finding the EM counterpart of the GW source, as deduced from the GW observation. Note that this means the maximum probability of imaging the EM counterpart in the current work is capped at approximately 95\%
, though this requirement can easily be relaxed.
  \item \textit{Tile}: An area of the sky covered by a single image. The shape of the tile matches the field of view of the telescope, and is assumed to be rectangular in our algorithms. Probability covered by any tile is calculated by convolving the localization PDF with the tile footprint (for instance, by adding the probabilities of all \texttt{HEALPIX} pixels inside the tile).
 There are a total of $N$ tiles to be imaged --- the arrangement of these tiles is discussed later. 
 We denote tiles by $I_1 \dots I_N$, short for ``Image''. For each tile $I_k$, $t_k$ is the last instant at which we can start imaging it and still complete the exposure before any part of the tile becomes unobservable.
  \item \textit{Tile rise/set}: Tile rise refers to the first instant at which the entire tile is visible. Tile setting refers to $t_k$ discussed above.
  \item \textit{Sunrise/set}: We use sunrise and sunset loosely to refer to astronomical twilight, when observing can commence or must end.
  \item \textit{Exposure time (\texp)}: We assume that each exposure takes \texp\ time, including readout and slew time. This is reasonable, as the slews will typically be small enough to be completed within readout time.
  \item \textit{Observation start time ($t_o$)}: Observations can start only at $t_o$. For observations to start, some part of the patch must be above the horizon, the telescope must be available (scheduling constraints), the sun must have set, and most importantly -- there must be a GW or transient trigger.
  \item \textit{Observation finish time ($t_f$)}: Observations end at $t_f$, either because the entire patch sets (goes below observable altitude), or the sun rises (or morning twilight), or the observing window is over (scheduling constraints). Note that patches can sometimes be multi--modal: there might be multiple disconnected ``islands'' that cover the 95\% confidence region. In such cases, the algorithm is run till \textit{every point} in the patch has either been observed, or has set. For simplicity, in this work we assume that the interval $t_f - t_o$ is an integral multiple of \texp.
  \item \textit{Imaging windows}: The observing interval ($t_o$ to $t_f$) is divided into ``Imaging windows'' $W_j$ of duration \texp\ each. There are a total of $M$ such imaging windows, where $M = \lfloor \frac{t_f - t_o}{\texp} \rfloor$. The $j^{\rm th}$ window starts at time $t_o + (j-1) \texp$ and ends at $t_o + j \texp$. For simplicity, we say that tiles with setting time $t_k$ in this range ``belong to'' the window $W_j$.
\end{itemize}

\subsection{Greedy Array (\grar)}\label{subsec:GrAr}
First we discuss the intuitive greedy algorithm. We superpose a grid on the sky localization patch to divide it into tiles shaped like the telescope field of view. Note that we have 3 free parameters in selecting the first tile of such a grid: the right ascension, declination and position angle (PA)\footnote{For ongoing synoptic surveys, the grid may be pre--defined for ensuring availability of reference images. This does not alter our algorithms.}. We set the PA to 0\degr, such that one side of all tiles is aligned with the celestial equator. We then convolve the footprint of a tile with the localization PDF to calculate the probability covered by a tile placed on each point in the patch. We select right ascension and declination such that first tile $I_1$ is centered on the point with the highest probability of finding the source. Using $I_1$ as reference, we create a column of tiles along the North--South direction by lining up other tiles with it. For each tile in this column, we construct rows by butting tiles edge to edge at a fixed declination. Note that as we move east or west from the central column, tiles from successive rows no longer line up as right ascension lines converge towards the pole.

The greedy algorithm is straightforward: in each imaging window $W_j$, we select the highest probability tile that is visible during the entire window (rises before $t_o + (j-1)\texp$ and sets after $t_o + j\texp$) and has not already been observed in a previous window. Since we have divided the patch into an array of tiles, we call this the Greedy Array algorithm (\grar). This algorithm naturally takes care of new tiles rising over time: if a just--risen tile covers higher probability than tiles that were already visible, it will be selected for observation. However, the algorithm cannot account for the setting of tiles: medium priority tiles may set while high priority (high probability) tiles are being observed, leaving no opportunity to observe the medium priority tiles later in that observing run.

As the ideal case, we also consider a greedy algorithm with no visibility constraints: for instance a space telescope that is (unrealistically) unaffected by solar or lunar constraints. In that case, we would observe the entire patch, i.e., all tiles, in order of probability covered. 
We call this algorithm ``Space Greedy'' (\spgr), and use it to determine an upper limit on the performance of any algorithm. Due to practical limitations, like the passage of day and night, rising and setting of the source, etc., any ground--based method will do worse than \spgr. \citet{sps12} observe that on an average, introducing horizon-- and sun--based visibility constraints decreases probability of imaging the counterpart by a factor of $\sim$3.

\begin{figure}[thbp]
\begin{center}
\includegraphics[width=\columnwidth]{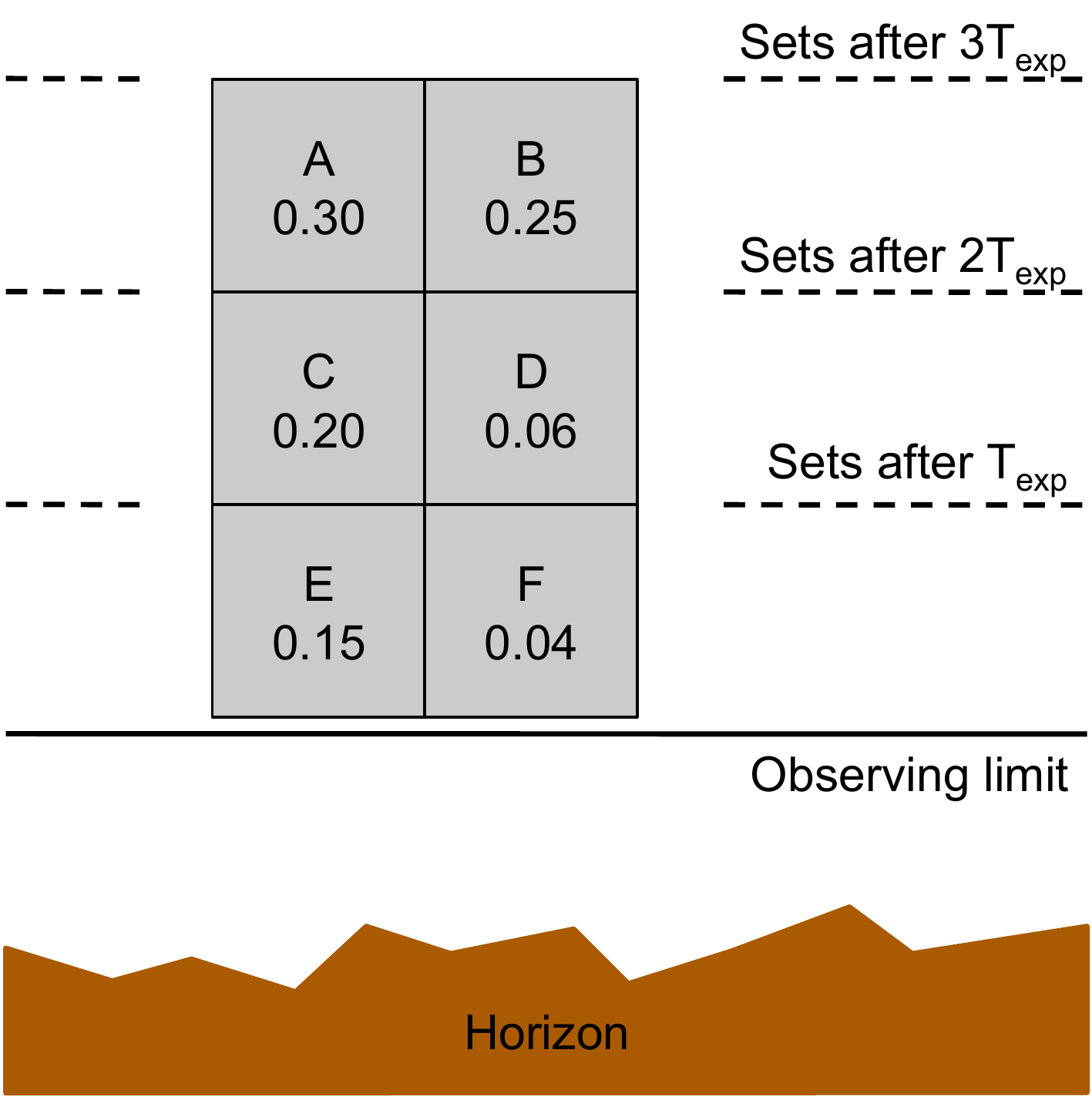}
\caption{A simplified $2\degr\times3\degr$ rectangular localization patch is covered by six $1\degr\times1\degr$ telescope images. The default greedy algorithm (\grar) generates the observation sequence \field{A}, \field{B}. The setting array algorithm (\sear) improves over this by generating the order \field{C}, \field{A}, \field{B}. This sequence is further optimized by \opar\ to observe higher probability tiles first, yielding the final sequence \field{A}, \field{C}, \field{B}.}
\label{fig:compare_all_ar}
\end{center}
\end{figure}

\subsection{Setting Array (\sear)}\label{subsec:SeAr}

\begin{table*}[!htbp]
\caption{Walkthrough of the SeAr algorithm\label{tab:searwalkthrough}}
\begin{center}
\begin{tabular}{|c|c|l|l|p{4cm}|l|}\hline
\textit{Iteration} & Tiles in this window & Candidate set & Past selection & Action & Final selection \\
$i$ &  & $\mathbb{C}_i$ & $\mathbb{S}_{i-1}$ & & $\mathbb{S}_i$ \\ \hline
1 & $\field{E}_{0.15}$, $\field{F}_{0.04}$ & \{ $\field{E}_{0.15}$ \} & \{ \} & Select highest probability tiles & \{ $\field{E}_{0.15}$ \} \\ \hline
2 & $\field{C}_{0.20}$, $\field{D}_{0.06}$ & \{ $\field{C}_{0.20}$, $\field{D}_{0.06}$ \} & \{ $\field{E}_{0.15}$ \} & $p_C > p_E > p_D$: retain \field{E}, reject candidate \field{D} & \{ $\field{E}_{0.15}$, $\field{C}_{0.20}$ \}  \\ \hline
3 & $\field{A}_{0.30}$, $\field{B}_{0.25}$ & \{ $\field{A}_{0.30}$, $\field{B}_{0.25}$ \}  & \{ $\field{E}_{0.15}$, $\field{C}_{0.20}$ \} & Both candidates are superior to past selection. Reject lowest probability tile \field{E}. & \{ $\field{C}_{0.20}$, $\field{A}_{0.30}$, $\field{B}_{0.25}$ \} \\ \hline
\end{tabular}
\end{center}
\tablecomments{See Figure~\ref{fig:compare_all_ar} for details. The subscript on a tile label denotes the probability that the true counterpart is located within that tile.}
\end{table*}

The greedy algorithm can run into problems when the patch is about to set. Consider a highly simplified patch shown in Figure~\ref{fig:compare_all_ar}. We have a six square degree rectangular patch, which can be imaged by six $1\degr \times 1\degr$ tiles, labeled \field{A}$\dots$\field{F}. Tiles  \field{E} and \field{F} are visible only for a duration \texp\ from the start of observations ($t_o$). Tiles \field{C} and \field{D} set at 2\texp, and tiles \field{A} and \field{B} set last at $t_o + 3\texp$. The greedy algorithm proposes that we observe tile \field{A} first. After obtaining a \texp\ minute exposure, tiles \field{E} and \field{F} are no longer visible. It therefore proposes that we observe the next highest probability tile: \field{B}. During that exposure, tiles \field{C} and \field{D} also set. The total probability covered, i.e. the probability of finding a counterpart in these observations is $p_{\rm GrAr} = p_A + p_B = 0.30 + 0.25 = 0.55$. For this simple case, we can easily devise some better strategies. For instance, we can observe tile \field{C} first, followed by tiles \field{A} and \field{B}, which are visible for a longer duration. In this new strategy, the total probability covered would be $p_C + p_A + p_B = 0.20 + 0.30 + 0.25 = 0.75$. 

We generalize this method, calling it the ``Setting Array'' algorithm (\sear).
It is conceptually simpler to examine the setting problem backward in time. Using terms from \S\ref{subsec:nom}, let us consider $M$ imaging windows $W_1 \dots W_M$, spanning the time range from $t_o$ to $t_f$. We remind the readers that $t_f$ can correspond to the last tile setting, or end of available observing time, or morning twilight. At the start of the last window $W_M$ (time = $t_f - \texp$), an observer should clearly observe the highest probability tile that is still visible. Going back to the imaging window $W_{M-1}$, more tiles which set in this window are also available for selection. So the observer will select the highest probability tile from all tiles visible at this instant. If there is an unobserved (or unscheduled) tile that belongs to $W_M$ that has a higher probability than tiles in $W_{M-1}$, that will be the tile observed in this window. We continue this way till we reach the imaging window $W_1$, starting at time $t_o$. This strategy gives high priority to the setting of tiles, and ensures that they are observed when they are above the horizon (or within more stringent observing limits, as the case may be).

The same algorithm can be implemented going forward in time. At time $t_o$, we initially select the highest probability tile (say $I_1$) that belongs to imaging window $W_1$, and add it to a ``selected'' set $\mathbb{S}_1$. By definition, this tile will not be visible in any further windows. For the next observation, we make a candidate set $\mathbb{C}_2$ with the two highest probability tiles from window $W_2$. Note that these tiles were also visible in window $W_1$: so if both of them cover higher probability than $I_1$, we can discard $I_1$ and image these two tiles in windows $W_1$ and $W_2$. Thus, the new selection $\mathbb{S}_2$ contains the 2 highest probability tiles from the set $\mathbb{S}_1 \cup \mathbb{C}_2$. We can inductively continue this process till the last imaging window. For the $k^{\rm th}$ iteration, the selected set $\mathbb{S}_k$ consists of the $k$ highest probability tiles from the set $\mathbb{S}_{k-1} \cup \mathbb{C}_k$. It is important to remember that these selected tiles are ordered: in each iteration, we add new tiles to the end of the set. This ensures that a tile added from the imaging window $W_k$ will be observed in $W_k$ or earlier. 
In Table~\ref{tab:searwalkthrough}, we apply this algorithm to the sample patch in Figure~\ref{fig:compare_all_ar} to recover the order \field{C}, \field{A}, \field{B} discussed at the start of this subsection.

There are some subtleties in this generalized algorithm. Firstly, it is possible that the localization patch is far from setting, so there may be no tiles that set in, say, the first $r$ windows. In such a case we select the highest probability tile from the next available window ($W_{r+1}$) for observation in $W_1$. Similar circumstances arise in intermediate steps if an imaging window $W_k$ has fewer than $k$ tiles. In such cases the candidate set $\mathbb{C}_k$ can use high probability tiles from the next imaging windows $W_{k+1}$, continuing to later windows as required. Lastly, the algorithm terminates at the last window $W_M$ or when no more tiles are visible. The latter case happens when the entire patch has set but more observing time is available.

\subsection{Optimized Array}\label{subsec:OpAr}

The Setting Array algorithm described above gives an observation sequence that maximizes the probability of detecting the EM counterpart of the GW source under the constraint that the patch will set in a given amount of time. However, the generated observing sequence is not unique. In practice, one prefers an algorithm that allows for the highest probability tiles to be observed first (like \grar), but still maximizes the overall probability covered (like \sear). We, therefore, seek a golden mean by reordering the \sear\ observing sequence to observe high probability tiles first, subject to the condition that each tile from the \sear\ observing sequence must be observed before it sets.

Let us reconsider the example discussed in Figure~\ref{fig:compare_all_ar} and Table~\ref{tab:searwalkthrough}. The final observing sequence proposed by the \grar\ algorithm would be \field{A}, \field{B}, with a total probability coverage $p_{\rm GrAr} = 0.30 + 0.25 = 0.55$. \sear\ improved on this by proposing the observing sequence \field{C}, \field{A}, \field{B} with a total probability coverage $p_{\rm GrAr} = 0.20 + 0.30 + 0.25 = 0.75$. Tile \field{C} does not set in the first imaging window, and can be observed the second window too. Thus, we can change the sequence to \field{A}, \field{C}, \field{B}: retaining the tiles selected by \sear, but reordering it to make faster gains on the total probability covered. In order to observe high probability tiles earlier, our guiding principle is to push low probability tiles to as late an observing window as possible.

We can formalize this algorithm as follows:
\begin{enumerate}
\item Use the \sear\ algorithm to create an observing sequence consisting of tiles $I_1 \dots I_M$ in imaging windows $W_1 \dots W_M$. The probability of finding the counterpart in tile $I_i$ is $p_i$.
\item Select the lowest probability tile, say $I_l$, and move it to the latest imaging window ($W_k$) where it is still visible. $W_k$ is now marked as ``fixed'', and the observation in this window cannot be changed further.
\item The \sear\ algorithm guarantees that tile $I_i$ is visible at least until window $W_i$. Thus, when we move a tile $I_l$ from current window $W_l$ to the last possible imaging window $W_k$, it always satisfies the condition $k \geq l$
\item Tiles originally occupying non--fixed windows in the range $W_{l+1} \dots W_k$ are moved to the earlier windows $W_l \dots W_{k-1}$.
\item Repeat the procedure with the next higher probability tile.
\end{enumerate}
The end result of this procedure is that high probability tiles move progressively towards earlier observing windows, giving a final sequence that prioritizes probability without losing any tiles to visibility constraints. 

\begin{figure}[htbp]
\begin{center}
\includegraphics[width=0.9\columnwidth]{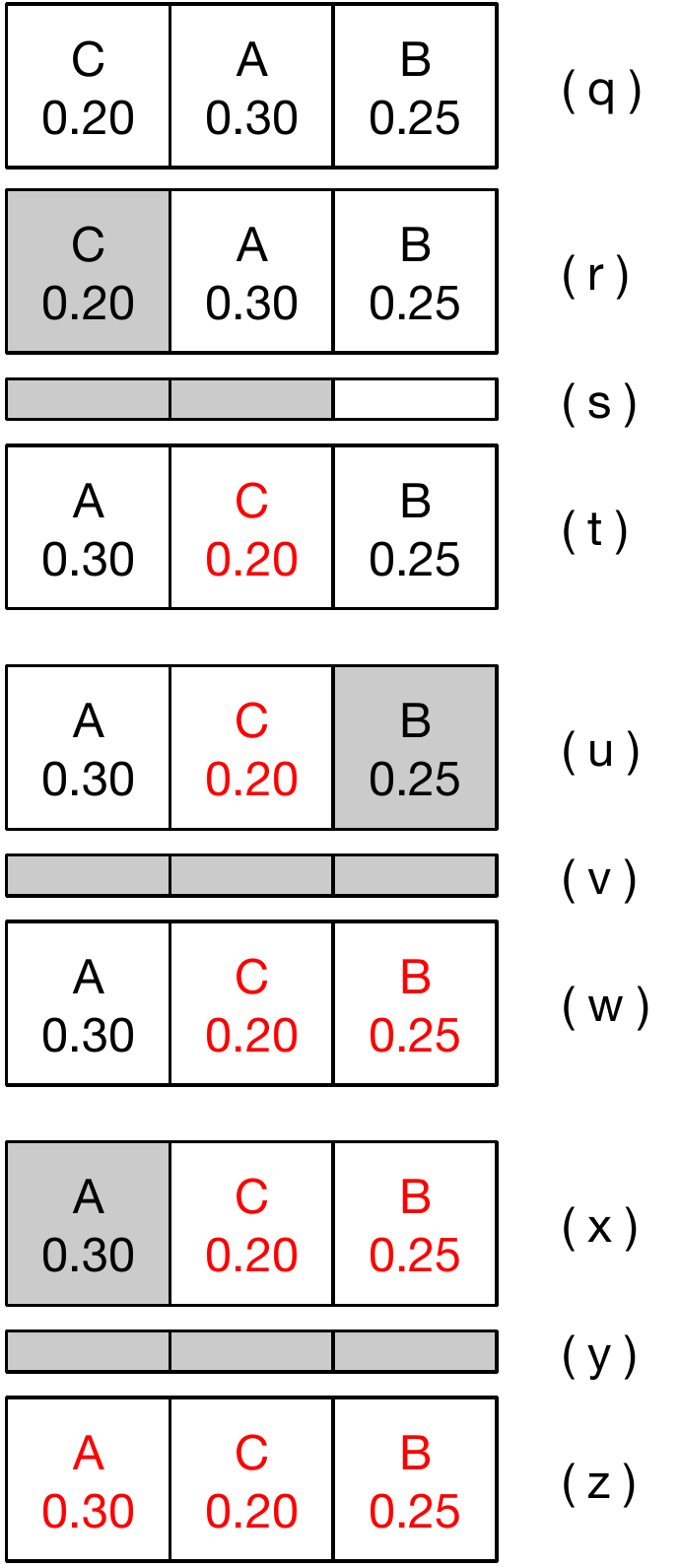}
\caption{Calculating an \opar\ observing sequence from a \sear\ sequence. We consider the same example as \S\ref{subsec:SeAr}. Capital letters denote tile names, and the numbers show the probability of containing the EM counterpart. Panel (q) shows the \sear\ observing sequence: tiles \field{C}, \field{A}, \field{B}. We select the lowest probability tile, \field{C} (panel r). Its visibility plot (shaded frames in panel s) shows that it can be moved to the second window. We move it to the second window, slide other tiles to the left, and lock it (red tile, panel t). Then we consider the second-lowest priority tile: \field{B} (panel u). \field{B} is visible in all imaging windows (panel v) and is already in the last observing slot, so we simply lock it in place (panel w). Finally, we consider the last tile, \field{A} (panel x). This tile cannot be moved as all other windows are locked. Panel y verifies that the tile is indeed visible in the only available imaging window, $W_1$. Panel z shows the final observing sequence for the Optimized Array algorithm: tiles \field{A}, \field{C}, \field{B}.}
\label{fig:OpAr_walkthrough}
\end{center}
\end{figure}

We call this algorithm the ``Optimized Array'' algorithm (\opar). Although \opar\ covers the same total probability as \sear, it observes higher priority tiles as early as possible. This proves to be advantageous in some cases. One such case is when the counterpart is likely to fade fast. \opar\ also performs slightly better than \sear\ when rising tiles are taken into account, as discussed in the next subsection. In Figure~\ref{fig:OpAr_walkthrough}, we apply this algorithm to our sample patch to resort the \sear\ observing sequence, and recover the \field{A}, \field{C}, \field{B} sequence discussed at the start of this subsection.


\subsection{Rising patch}\label{subsec:rising}
The greedy algorithm is well suited for patches that are always visible, or rise as observations progress; but runs into trouble with setting patches. Our proposed \sear\ and \opar\ algorithms address this shortcoming. However, some patches consist of long arcs, where some parts of the patch set during the night while other parts rise during the same period of time.
We adapt \sear\ and \opar\ to this situation by the following prescription:
\begin{enumerate}
  \item Select the part of the patch that is observable at time $T$.
  \item Run \opar\ (or \sear) on the patch, and select the tile $I$ that should be observed in the next imaging window $W_k$.
  \item \label{step:advance} Advance time by $\texp$. Some more tiles from the patch may rise in this interval.
  \item Run \opar\ (or \sear) on all tiles that are visible now but have not been observed yet. The first tile selected by the algorithm should be observed in the next imaging window $W_{k+1}$.
  \item Repeat from step~\ref{step:advance} until all tiles are observed, or till observing finish time ($t_f$).
\end{enumerate}
Hence, on each iteration we run the full algorithm on the visible patch, but schedule only the first tile as the next observation. If no tiles rise before the next iteration, one can see that the tile sequence will remain the same as a single iteration of \sear\ (\S\ref{subsec:SeAr}) or \opar\ (\S\ref{subsec:OpAr}). On the other hand, if important tiles rise during the observation, the algorithms will incorporate them in the optimal scheduling. A trivial example is the case where the highest probability tile rises just for the last observing window $W_m$. Following the \sear\ algorithm, this tile enters the candidate set $\mathbb{C}_j$. Being the highest probability tile, it trumps all tiles from the past selection and gets scheduled for the final observation.

As an aside, we note that \sear\ and \opar\ as defined in \S\ref{subsec:SeAr} and \S\ref{subsec:OpAr} respectively, always have the same set of tiles for final observation. However, after applying this prescription, in some cases, the final selected set of tiles may differ slightly (\S\ref{subsec:array}, Figure~\ref{fig:patch_ar}).

\begin{figure}[thbp]
\begin{center}
\includegraphics[width=\columnwidth]{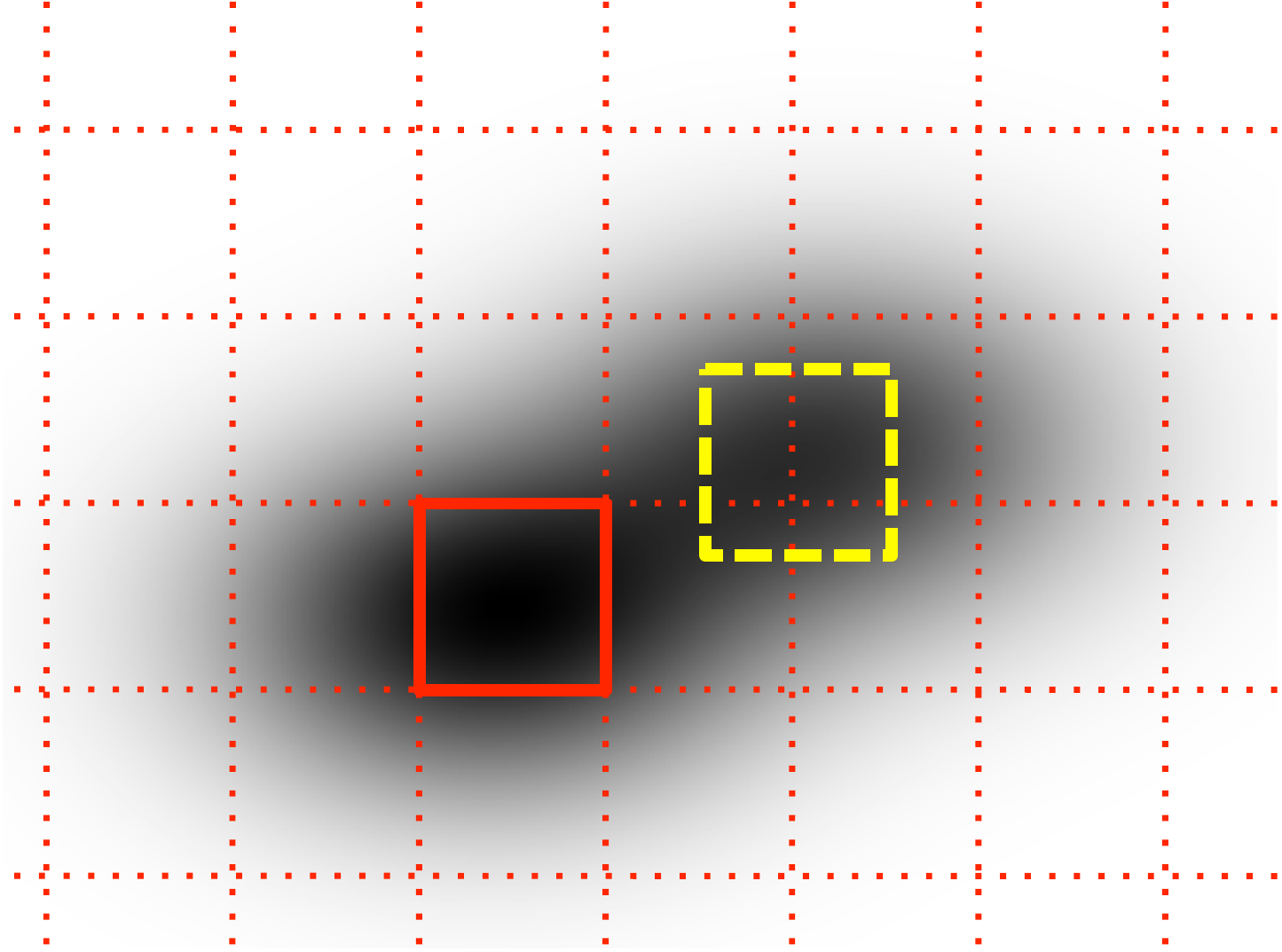}
\caption{The need for Independent tiling. We consider a small patch that is about to set soon. If only two observations are possible, then observing the thick red and yellow tiles maximizes the probability coverage. Note that the yellow tile is placed independently of the grid created by the placement of the first thick red tile.}
\label{fig:GrIn_example}
\end{center}
\end{figure}

\subsection{To array or not to array}\label{subsec:independent}

Sometimes, based on the geometry of the patch and the size of the observing tiles, an array (grid) of tiles on the sky may be too constraining. Let us consider a simplistic case in Figure~\ref{fig:GrIn_example} where the PDF has two local maxima. Following \S\ref{subsec:GrAr}, the first tile (thick red border) is placed maximize the probability of detecting the counterpart in the first image. This sets up the array (dotted red lines) that is used to select subsequent tiles. However, if only one more image is to be obtained, then better results can be obtained by placing the next tile (dashed yellow square) independent of the existing array.

To investigate if such free placement of tiles is effective, we propose three ``Independent'' algorithms that correspond to the three ``Array'' algorithms discussed above (see Table~\ref{tab:algos}). The first is the greedy algorithm with independent placement of tiles (``Greedy Independent'', or \grin). In each imaging window $W_k$, we select the tile that has the highest probability of containing the EM counterpart. We then set the probability of that part of the patch to zero\footnote{In a strict sense, the resultant distribution cannot be called a PDF as its sum over the entire sky is no longer unity. However, the value at each point other than the zeroed out region is still the probability density of finding the EM counterpart at that location.}, and calculate the next best location for the next image, using whatever patch is visible at that time. The zeroing out is based on the assumptions in \S\ref{subsec:nom} that repeat observations do not increase the probability of finding the counterpart. 

Allowing each tile to be placed independently of other tiles is a problem for the \sear\ and \opar\ algorithms, which assume that the probabilities and setting times for all $M$ tiles are known at the start while computing the observing sequence. This cannot be done if the tile positions are not known. 
In practice, neighboring tiles have similar setting times and are seen to have similar probability of containing the EM counterpart, in particular when the tile area is much smaller than the patch area.
As a result, we heuristically extend our Setting Array and Optimized Array algorithms as follows: 
\begin{enumerate}
\item Calculate the full sequencing for \sear\ (\S\ref{subsec:SeAr}) or \opar\ (\S\ref{subsec:OpAr}).
\item Observe the first tile from the final sequence ($I_1$). \label{item:firstobs}
\item Update the localization patch to exclude the tile already observed, by setting the PDF to zero in that area.
\item For the $j^{\rm th}$ iteration, update $t_o \rightarrow t_o + (j-1)\texp$ and rerun the algorithm to calculate the next complete observing sequence. Select first tile as $I_j$.
\item Repeat until all tiles are observed or the observing period ends (time = $t_f$).
\end{enumerate}
The location of tiles of each iteration is independent of the array created in the previous iteration. Hence, we name these two algorithms as ``Setting Independent'' (\sein) and ``Optimized Independent'' (\opin) respectively. 

This gives us a total of seven algorithms: the completely unrestricted ``Space Greedy'' (SpGr) discussed in \S\ref{subsec:GrAr}, and the six algorithms listed in Table~\ref{tab:algos}. We compare the performance of these algorithms in the next section.

\begin{table}[hbtp]
\caption{List of algorithms\label{tab:algos}}
\begin{center}
\begin{tabular}{|l|c|c|c|}\hline
\textit{Selection} $\longrightarrow$ & Greedy & Setting & Optimized \\
\textit{Pattern} $\downarrow$ & \textit{(Gr)} & \textit{(Se)} & \textit{(Op)} \\\hline
Array \textit{(Ar)} & \grar & \sear & \opar \\\hline
Independent \textit{(In)} & \grin & \sein & \opin \\\hline
\end{tabular}
\tablecomments{Apart from these six algorithms, we also consider the ``Space Greedy'' (\spgr) algorithm in some of our simulations.}
\end{center}
\end{table}

\section{Simulations}\label{sec:simulations}

\begin{figure*}[thbp]
  \centering
    \includegraphics[width=0.53\textwidth]{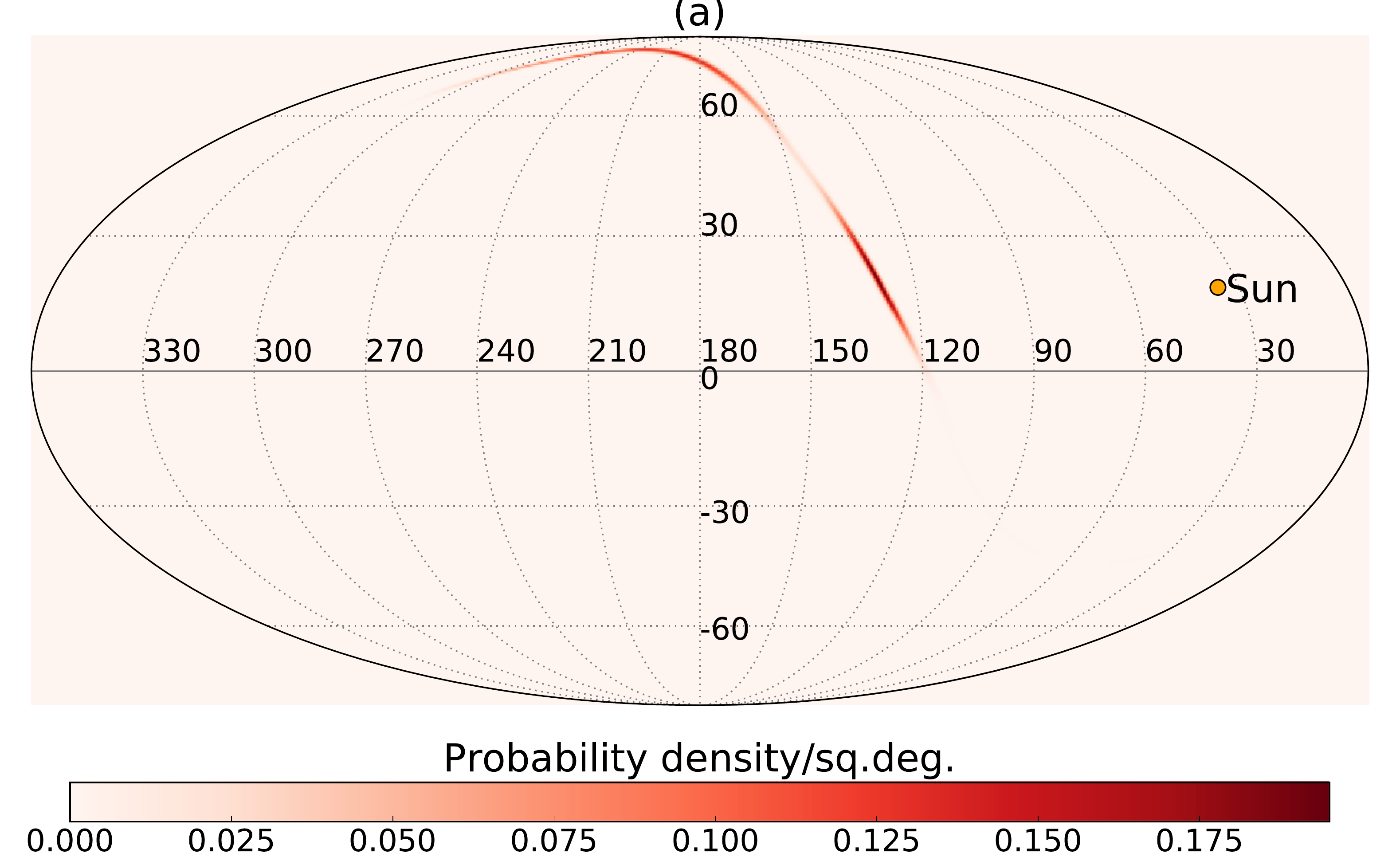}\hspace*{0.5cm}
   \includegraphics[trim=2cm 0 3cm 1cm,clip=true,width=0.41\textwidth]{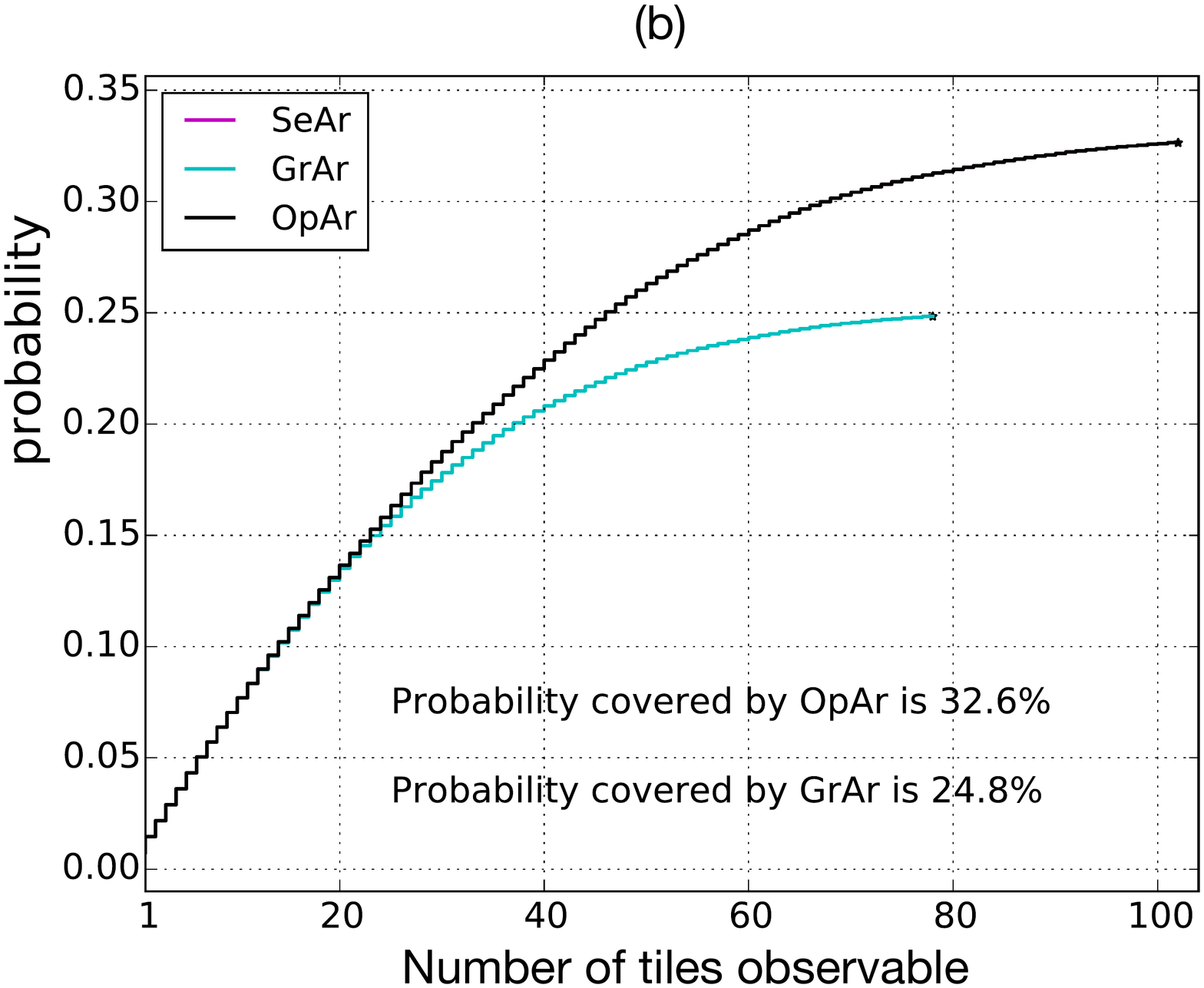}
  \caption{Comparing \sear, \opar\ and \grar. Panel (a) shows the all--sky probability density function of the patch in equatorial coordinates, with a 95\% containment area of 305~\sqd\ The northern island contains about 40\% of the probability and is visible from IGO, while the more prominent patch near the center is too close to the sun. Panel (b) compares the probability coverage of the three methods, if the observations are terminated after $n$ images. The X--axis is the number of tiles, and the Y--axis shows the total probability covered by Array methods when observations end. The \opar\ (black) and \sear\ (magenta) curves are coincident in this plot. The performance of both these methods is better than, or at least equal to, \grar.
    \label{fig:compare_ar_envelope}}
\end{figure*}

\begin{figure*}[!hbp]
  \centering
    \includegraphics[width=0.5\textwidth]{305_patch349_mol2.pdf}\hspace*{0.5cm}
    \includegraphics[width=0.44\textwidth]{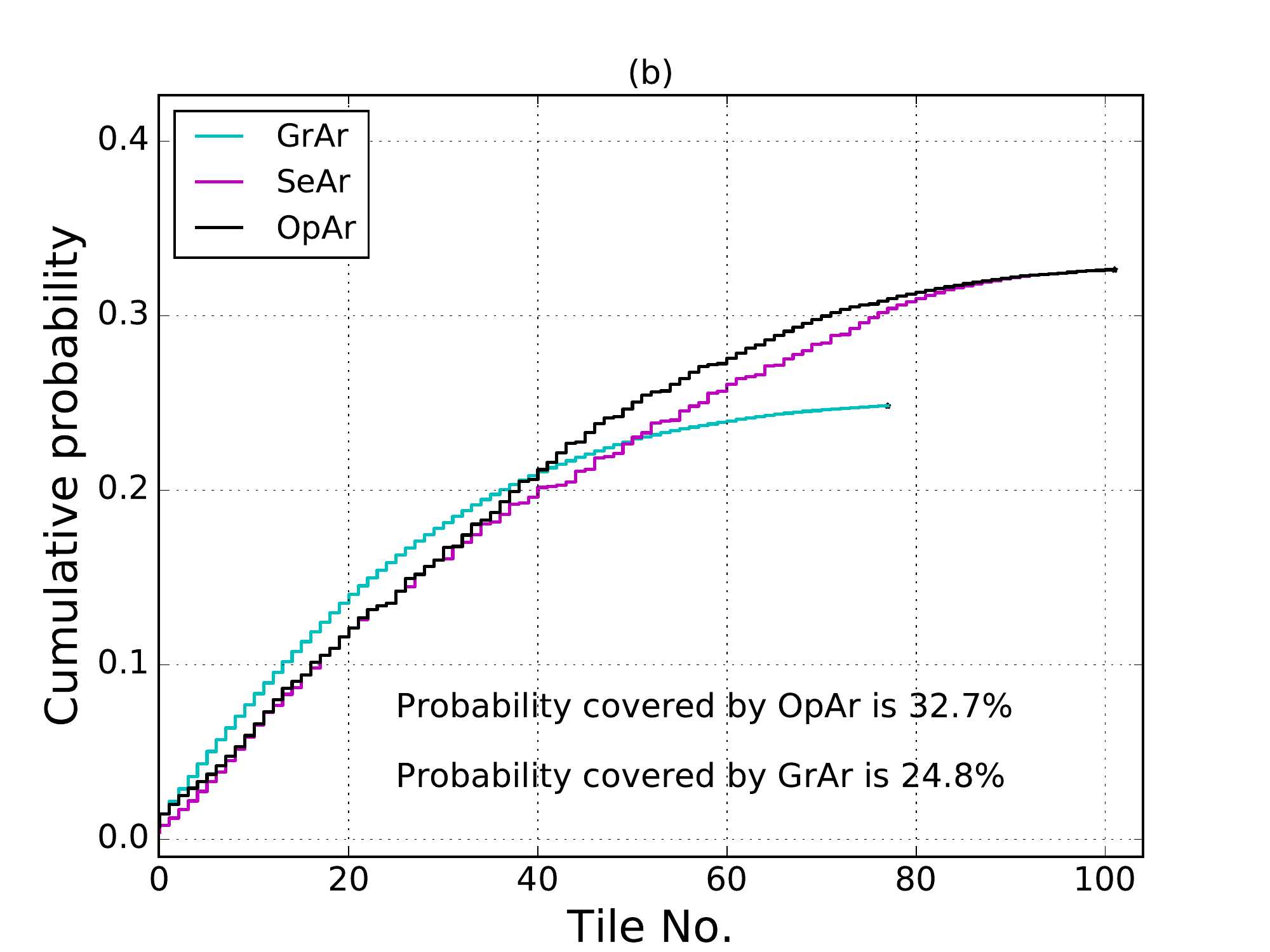}
    \includegraphics[width=1.0\textwidth]{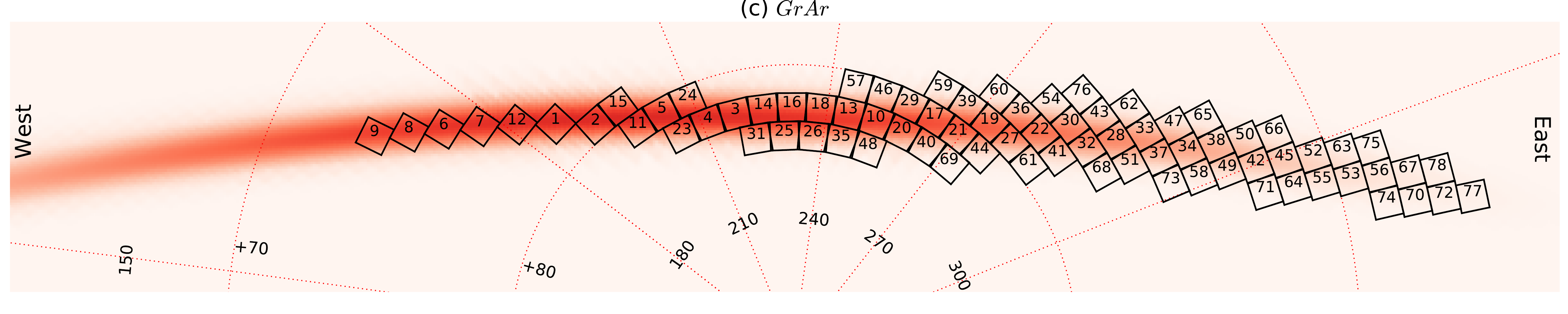}
    \includegraphics[width=1.0\textwidth]{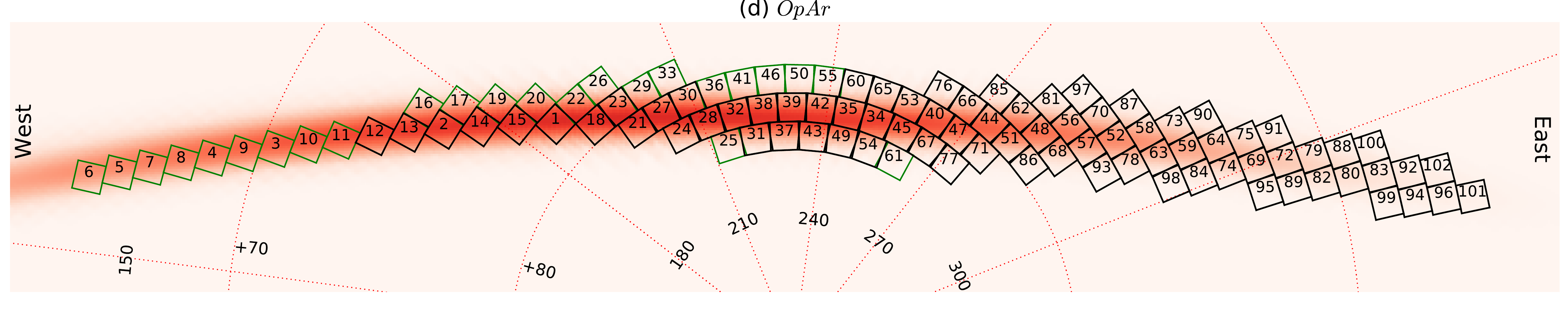}
  \caption{Comparing \sear, \opar\ and \grar. Panel (a) shows the all--sky probability density function of the patch in equatorial coordinates, with a 95\% containment area of 305~\sqd\ The northern island contains about 40\% of the probability and is visible from IGO, while the more prominent patch near the center is too close to the sun. Panel (b) compares the probability coverage of the three methods after $n$ images, \textit{given that end of observations is determined only by patch setting and sunrise}. The X--axis is the number of tiles, and the Y--axis shows the cumulative probability covered by Array methods. Panel (c): the \grar\ algorithm covers $24.8\%$ probability in 78 tiles. Numbers in the tiles denote observation order. Approximate East and West directions are indicated: \grar\ misses the medium probability regions to the west of the high probability center. Panel (d): \opar\ and \sear\ cover $32.7\%$ probability by imaging 102 tiles. Apart from the tiles calculated by \grar\ (shown in black), \opar\ also selects some western tiles (shown in green) for early observations, boosting the overall probability coverage.
    \label{fig:compare_ar}}
\end{figure*}

We tested the performance of these algorithms by simulating ground--based searches for counterparts to GW triggers. We assumed a telescope with a $1\degr \times 1\degr$ square field of view at the location of the IUCAA Girawali Observatory (IGO): longitude 19.08\degr~N, latitude 73.67\degr~E, and altitude 1000~m~\citep{gbd+02}. The observing limit is set to an altitude of 25\degr\ (where the airmass is 2.4; see the discussion in Sec.~\ref{subsec:airmass} below). We assumed that the exposure, readout and slew take a total of 300~sec for each tile. Our simulated observations start and stop when the sun is 12\degr\ below the western and eastern horizon, respectively. These  observations always end at the next sunrise after a trigger---so for triggers occurring shortly before sunrise, only a short observational window is used. We do not impose any lunar constraints in these simulations.

\cite{Singer_Localization_2014} have simulated the detection and localization of GW triggers using expected sensitivities for the first two years (2015--2016) of operation of advanced LIGO and VIRGO detectors. We simulate ground--based optical observations for all the 1609 binary neutron star merger event localizations that they provide. For some events, they provide localizations using two different algorithms --- in such cases, we treated both these as separate events in our follow--up simulations.
The 95\% containment areas for these patches range from tens to thousands of square degrees. We assume that GW detectors rapidly detect and verify a trigger, so that observations can start just ten minutes after the event. Of the 1609 patches, about one--third are not visible from IGO within 24~hours due to latitude constraints or proximity to the sun. Of the remaining 1054 patches, many are highly elongated and are only partially visible from IGO. Even in such cases, we compare the performance of our algorithms for the parts of the patch that are visible from IGO.

We implemented our algorithms using \texttt{python}. Sky localizations provided by \texttt{LVC} are in the \texttt{HEALPIX} format~\citep{ghb+05}, like files generated by \bay\footnote{\url{https://ligo-vcs.phys.uwm.edu/cgit/bayestar/tree/bayestar}}. Each \texttt{HEALPIX} pixel gives the probability that the true location of the GW source lies within that pixel. To calculate the probability that an image covers the true location of the source, we add up the probabilities of all \texttt{HEALPIX} pixels within that image. For efficiency, we add up the probability of all pixels whose centers are inside the image footprint. This introduces quantization errors for pixels along the image boundary, which lie only partially inside the image. In order to ensure that such quantization errors do not affect our results, we oversample the \texttt{HEALPIX} file to $N_{side} = 2048$, such that each pixel is $1.7^\prime$ on a side. For our fiducial 1~\sqd\ camera, this gives $\sim1220$ pixels per camera field (tile). In contrast, the usual files with $N_{side} = 512$ give only about seventy--six pixels per square degree on the sky, leaving the simulations susceptible to significant quantization errors.

\subsection{Array methods}\label{subsec:array}
First, we compare the performance of the array methods, where a single grid of tiles is superposed on the sky, and each method gives an ordered sequence of tiles to be observed. Along with the Greedy Array (\grar), Setting Array (\sear), and Optimized Array (\opar) methods, we also consider the Space Greedy (\spgr) method, which sets the upper bound on the performance of all other methods.

It is important to remember that the observations finish when any one of three conditions is met: either the entire patch sets (i.e., it goes below the observable altitude), or the sun rises (morning twilight), or the observing window is over (scheduling constraints). As a result, observing sequences for different algorithms may terminate at different times. In the example in Figure~\ref{fig:compare_all_ar}, the \grar\ observing sequence terminates after two images (\S\ref{subsec:GrAr}) while \sear\ and \opar\ terminate after three images. In these simulations, we also end \spgr\ when the \opar\ sequence terminates. 

The computation time for generating an observing sequence varies with the patch size. \opar\ involves the most computations, and is relatively slower. On a desktop computer (64-bit, 3.6~GHz, 8~GB RAM), \opar\ calculations take $\sim$42 seconds for a 500~\sqd\ patch, increasing to $\sim$3~minutes for patches with areas 7000--8000~\sqd

As an illustration, we show comparison between the performance of \opar\ and \grar\ for a 305~\sqd\ patch (Figure~\ref{fig:compare_ar_envelope}). The patch has two separate islands: the northern part containing $\sim 40\%$ probability of containing the counterpart is visible from IGO at night, while the other (covering $\sim 60\%$ of the probability) is hidden in daylight. The \grar\ algorithm provides a single observing sequence, which simply terminates when at $t_f$: when the patch sets or at sunrise. In contrast, \sear\ and \opar\ give observing sequences based on the number of observations possible. In each case, we see that \sear\ (magenta curve) and \opar\ (black curve, exactly coincident with the magenta curve) match or outperform \grar\ (cyan curve). For instance, if only sixty observations are possible, \grar\ covers a net probability of about 25\%, while \opar\ and \sear\ cover about 27\% probability of finding the counterpart. For reasons discussed in \S\ref{subsec:GrAr}, \grar\ stops observations at tile 78 as all visible areas have already been imaged. Even if more observing time were available, the \grar\ algorithm has no use for it. In contrast, \opar\ and \sear\ algorithms optimize the observing sequence between high--importance and early--setting tiles such that observations can continue to tile 102.

\begin{figure}[htbp]
\centering
    \includegraphics[trim=0cm 0 1cm 0,clip=true,width=\linewidth]{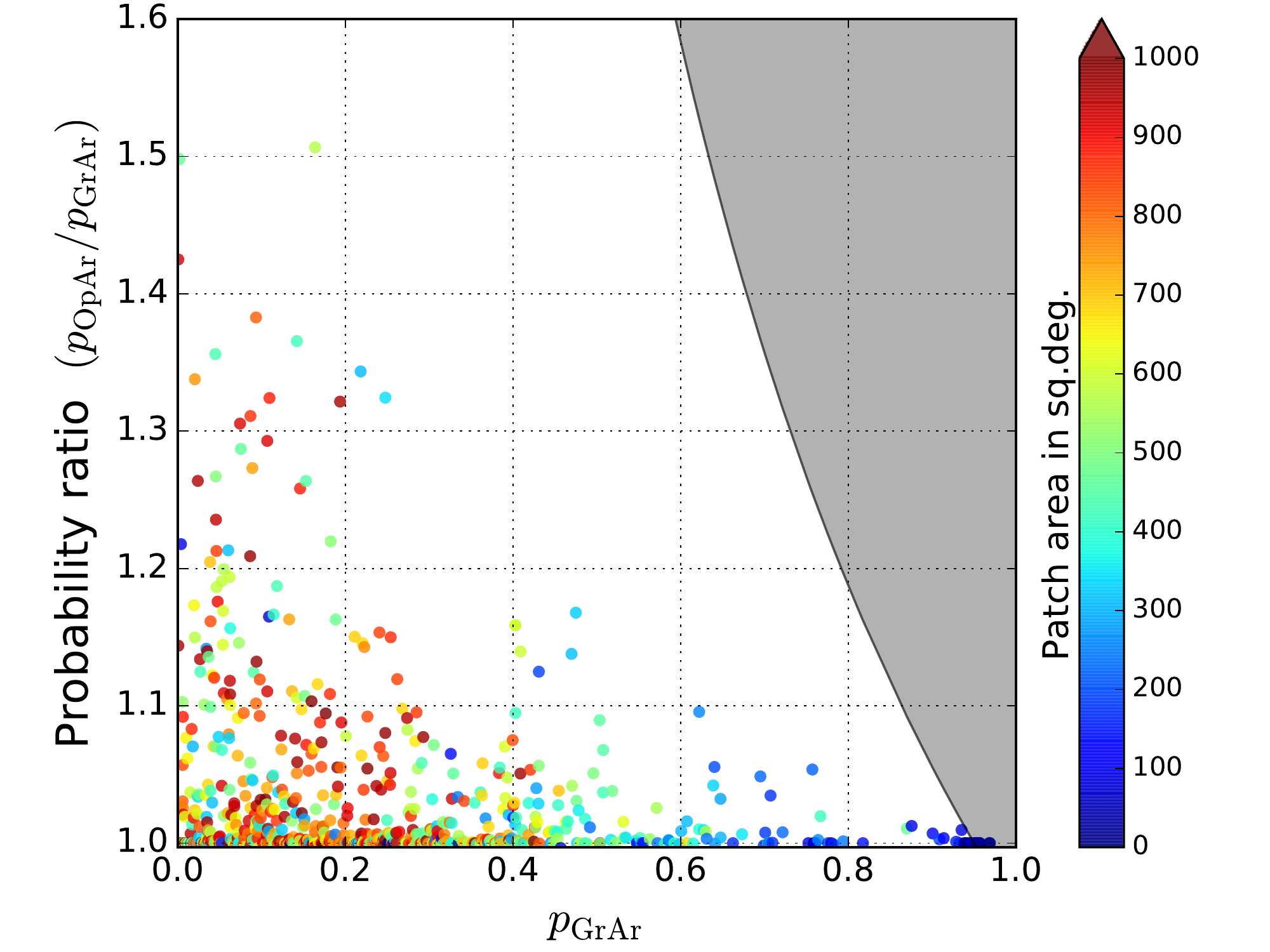}
    \caption{Comparing \opar\ with \grar. The X--axis is the total probability covered by \grar\ ($p_{\rm GrAr}$). The Y--axis shows the ratio of total probability covered by \opar\ to that covered by \grar. In the online color version, the color of the symbol shows the area of the 95\% patch in square degrees, with red denoting 1000~\sqd\ or larger patches. The shaded gray region on the right is excluded as $p_{\rm OpAr} > 0.95$ or $p_{\rm GrAr} > 0.95$. In all the 1054 patches we tested, \opar\ outperforms \grar. 
    \label{fig:results_ar}}
\end{figure}

\begin{figure}[htbp]
\centering
    \includegraphics[trim=0cm 0 1cm 0,clip=true,width=0.9\linewidth]{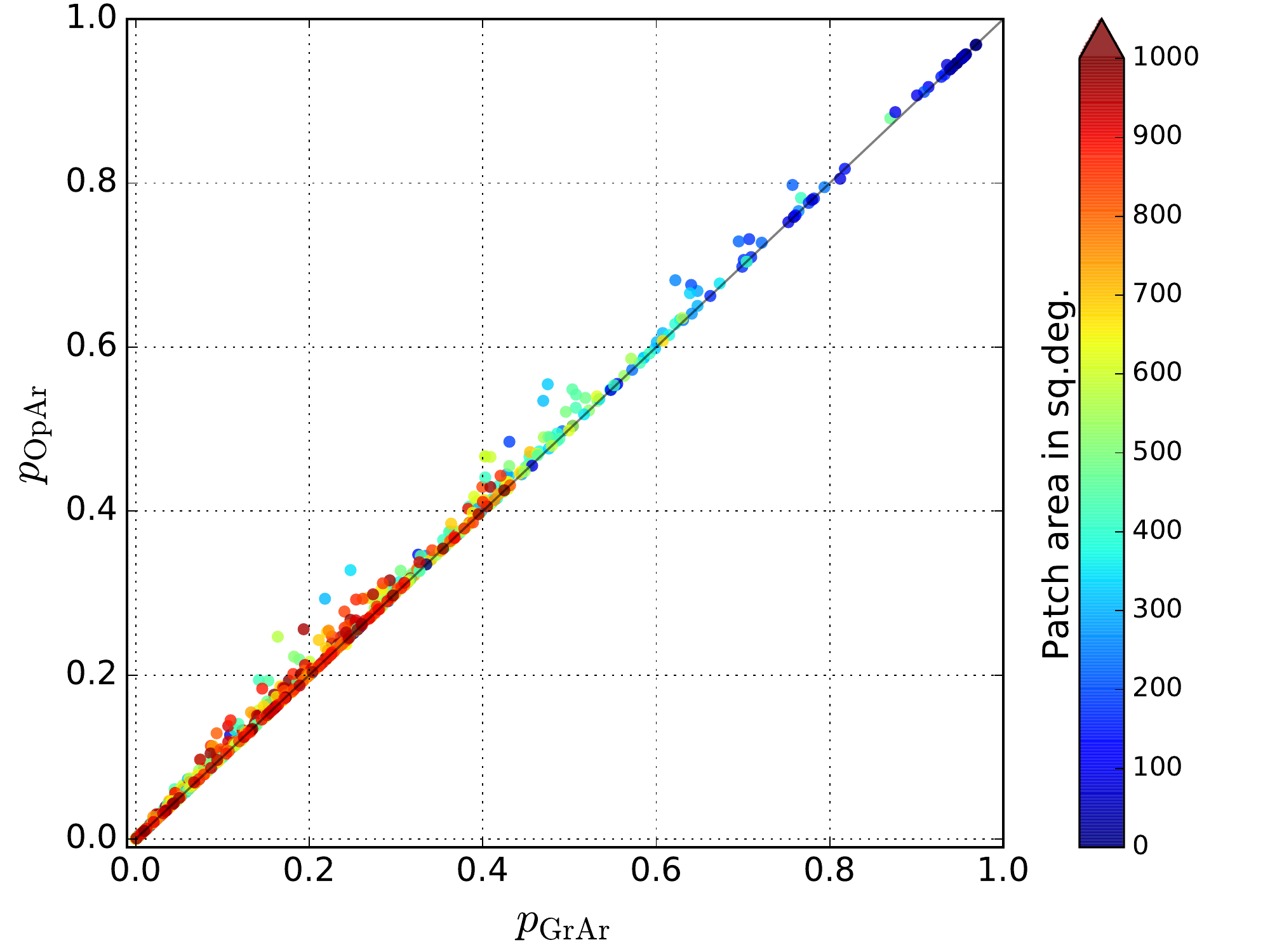}
    \caption{Comparing \opar\ with \grar. The X--axis and Y--axis are the total probability covered by \grar\ ($p_{\rm GrAr}$) and \opar\ ($p_{\rm OpAr}$) respectively. In the online color version, the color of the symbol shows the area of the 95\% patch in square degrees, with red denoting 1000~\sqd\ or larger patches. \label{fig:OpAr_GrAr}}
\end{figure}

\begin{figure*}[thbp]
\centering
    \includegraphics[trim=0.8cm 0 0.8cm 0,clip=true,width=\linewidth]{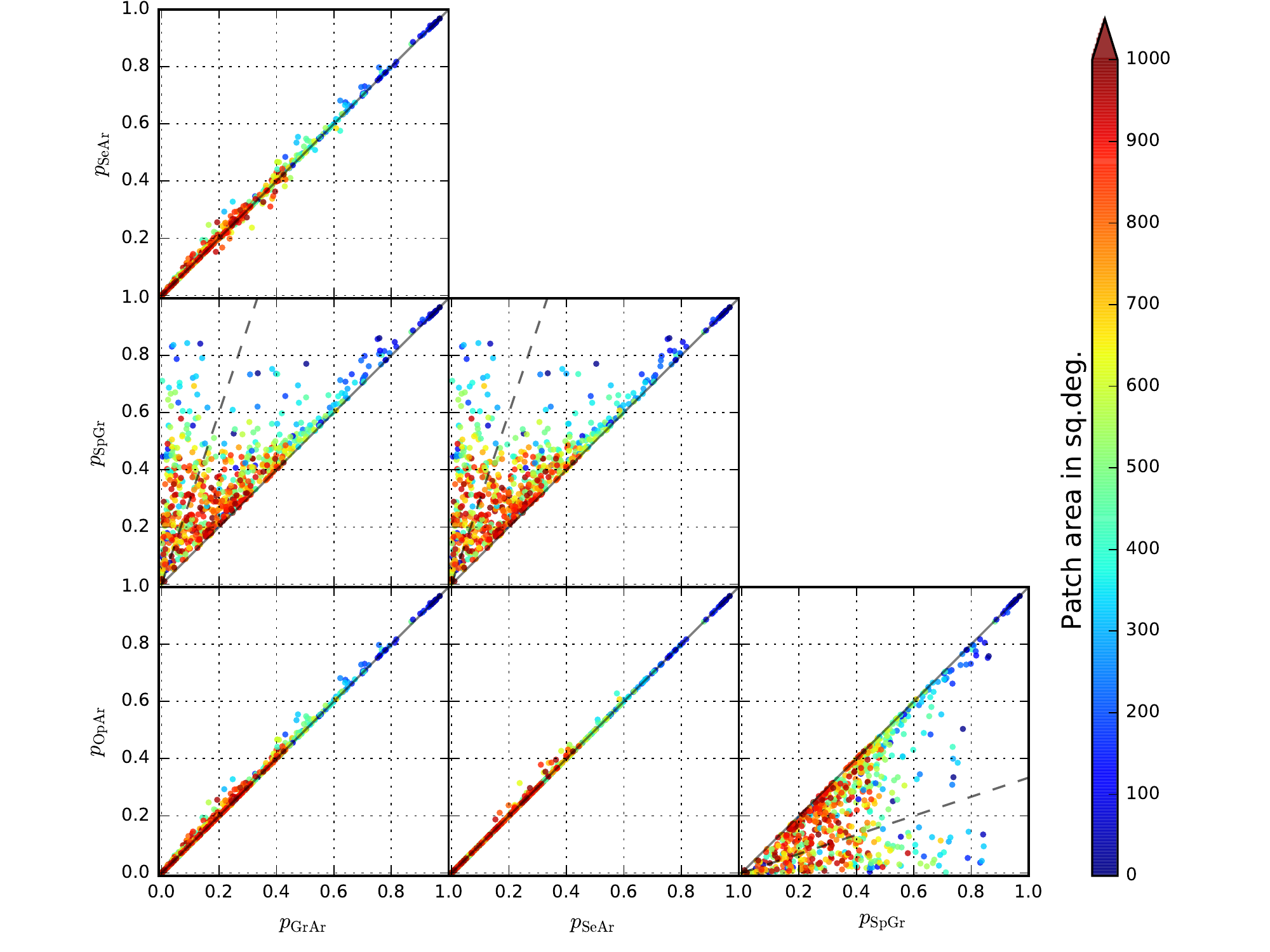}
    \caption{Comparing \opar, \sear, \grar\ and \spgr. In the online color version, the color of the symbol shows the area of the 95\% patch in square degrees, with red denoting 1000~\sqd  and larger patches. On an average, the completely unconstrained \spgr\ algorithm is expected to cover about thrice the probability as ground--based observations. This ratio is denoted by a dashed gray line in the three panels with \spgr\ on one of the axes. We see that \opar\ performs at least as well as, or better than, \grar. \label{fig:all_ar}}
\end{figure*}

Once the number of observations is calculated (from patch setting, sunrise etc.), \opar\ and \sear\ algorithms sequence tiles differently from \grar. This is illustrated in detail for the same patch in Figure~\ref{fig:compare_ar}. Owing to its sky location (\ref{fig:compare_ar}a), the patch sets well before sunrise --- so observations terminate when the entire patch has set. Panel (b) shows the cumulative probability covered by the three algorithms after $n$ images, \textit{given that end of observations is determined only by patch setting and sunrise}.
As expected, the greedy algorithm rapidly covers higher probability tiles first. However, the marginal gains become smaller as the high probability part of the patch becomes unobservable, and the cyan curve plateaus out around image 40. By image 78, all tiles that are still above the horizon have been imaged already (Figure~\ref{fig:compare_ar}c), and the total probability of having imaged the true source location is 0.25. In contrast, \sear\ and \opar\ start out slower, but do not hit a plateau till around image 80. By observing the western, early--setting tiles first, these algorithms enable 102 observations (Figure~\ref{fig:compare_ar}d) and cover higher overall probability: 0.33 as compared to 0.25 for \grar. The net probability covered by these two algorithms is the same, but \opar\ covers it faster than \sear. Even at the time when \grar\ observing terminates, \sear\ and \opar\ are performing better.

If practical considerations allow fewer images, say just 20, then the observation end time ($t_f$) can be changed while running the algorithm, to obtain an optimized sequence for the shorter time duration. While the \grar\ observing sequence remains unaltered by this, \sear\ and \opar\ output a new sequence that still covers a higher net probability compared to \grar.
\opar\ outperforms the basic \grar\ algorithm for all 1054 patches that we tested. Figure~\ref{fig:results_ar} shows the ratio of the probability covered by the two algorithms, as a function of the the baseline probability covered by \grar. In the online color version, the color coding shows 95\% area of the patch, in \sqd\ Note that by definition, a patch is a region with 95\% probability of containing the EM counterpart --- hence the maximum probability covered by any algorithm in these simulations is 0.95. The shaded gray region on the right is excluded as $p_{\rm OpAr} > 0.95$ or $p_{\rm GrAr} > 0.95$. It is seen that for those small patches that are covered well by \grar, the gain from \opar\ is marginal (right side). But for medium and large patches where \grar\ performance is poor, \opar\ can give a significant boost in the total probability covered. We can visualize this in another way, by plotting the total probability covered by two methods against each other (Figure~\ref{fig:OpAr_GrAr}). The X--axis and Y--axis are the total probability covered by \grar\ $(p_{\rm GrAr})$ and
\opar\ ($p_{\rm OpAr}$) respectively, and as before, the color coding shows the area of the patch. Each point in the plot represents the total probability covered by \grar\ and \opar\ for a single patch in one night. The black diagonal line denotes equal probability coverage for both algorithms. All points lie on or above the line, as \opar\ outperforms the probability coverage by \grar.

Lastly, Figure~\ref{fig:all_ar} compares the performances of all four array methods. As expected, \spgr\ outperforms the other algorithms, as it completely ignores visibility constraints. A counterintuitive feature in this plot is that $p_\sear$ and $p_\opar$ are not always equal. The algorithms as discussed in Sections~\ref{subsec:SeAr} and \ref{subsec:OpAr} assume that the entire patch is visible when observations start, and may eventually set. In practice, several patches will be visible partially when observations start, and more areas will be observable as time passes by. In our simulations, we apply a heuristic fix to such patches, as discussed in \S\ref{subsec:rising}. The under--performance of \sear\ as compared to \opar\ stems from the fact that \sear\ gives maximum importance to the setting time of a tile. Thus, it recommends observations of lower probability tiles that rise early, as compared to higher probability tiles that may be visible later. Since \opar\ observes higher probability tiles where possible, it covers more probability than \sear\ (as well as \grar) for all patches.

\begin{figure*}[!phtb]
\centering
    \includegraphics[width=0.51\textwidth]{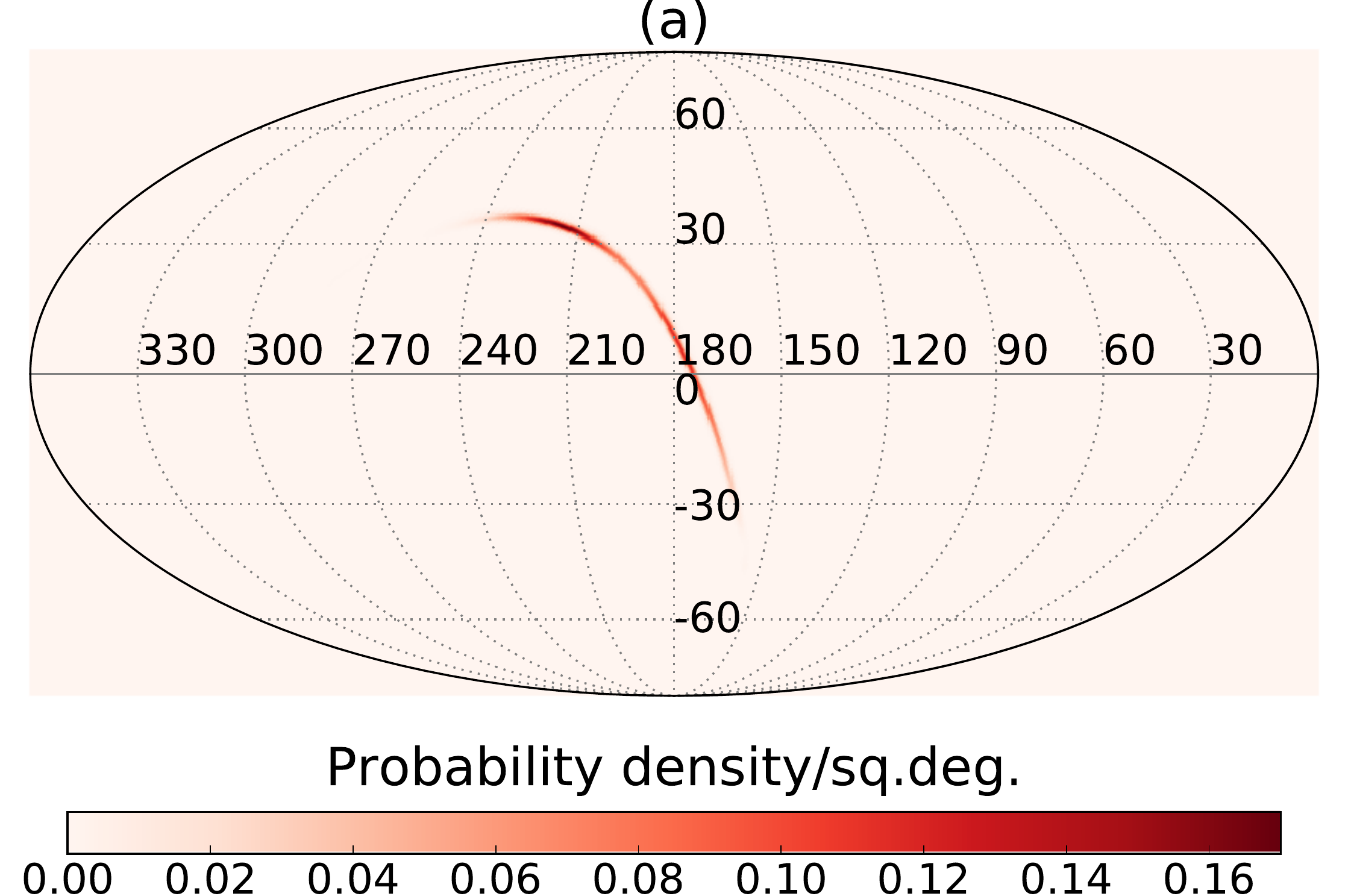}\hspace*{0.5cm}
    \includegraphics[width=0.45\textwidth]{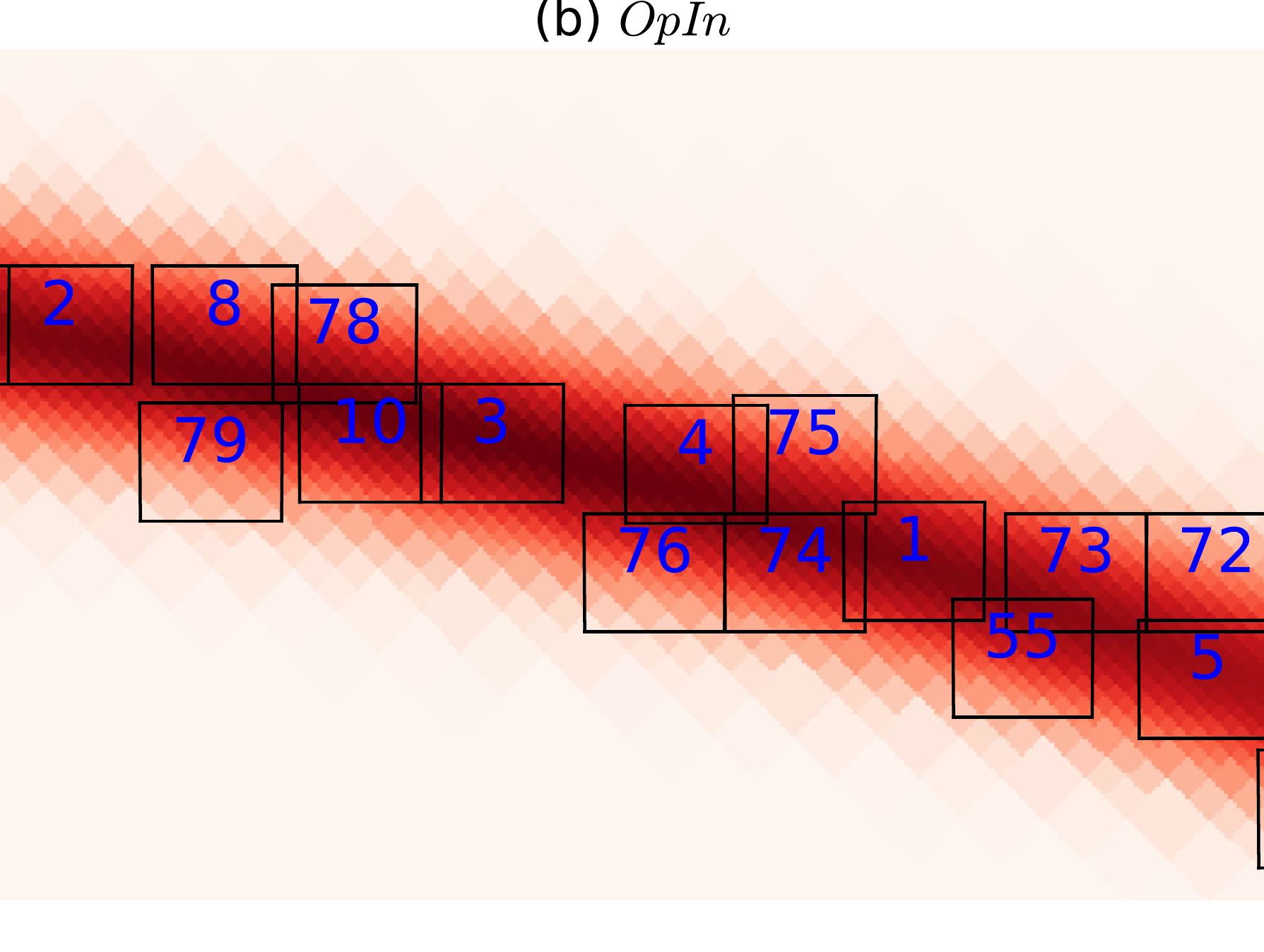}
    \includegraphics[trim=2cm 0 2cm 0cm,clip=true,width=0.47\textwidth]{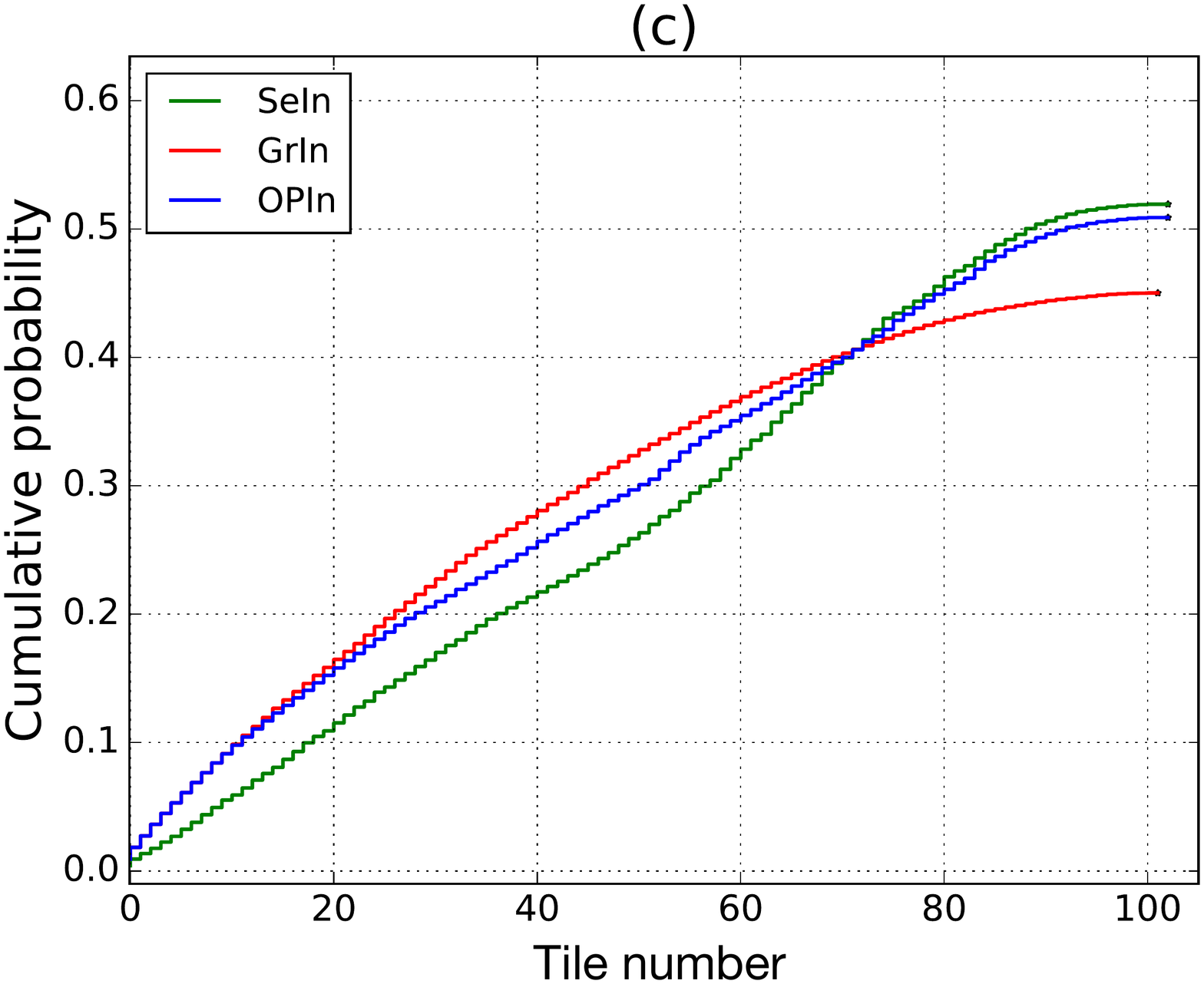}\hspace*{0.5cm}
    \includegraphics[trim=2cm 0 2cm 0cm,clip=true,width=0.47\textwidth]{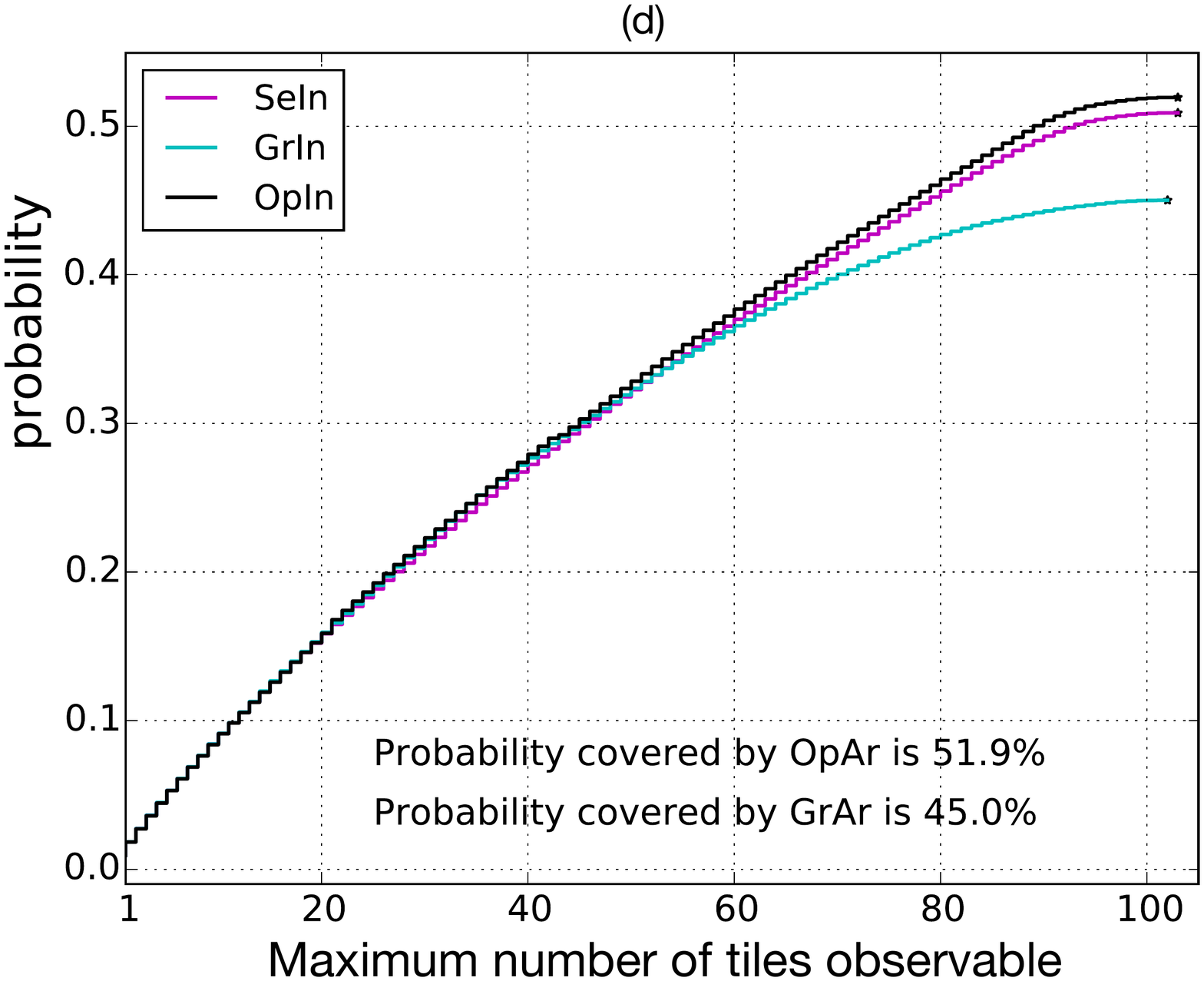}
    \caption{Comparison of the three independent methods. Panel (a) shows the Molleweide projection (in equatorial coordinates) of a long arc shaped patch of 313\sqd, color coded by probability density. Panel (b) is a zoom--in on a part of the observing sequence calculated by \opin, where each tile is numbered by its \opin\ observing order. Note how tiles overlap with each other, and late observations (78, 79\dots\ etc.) come back to the central high probability regions. Panel (c) shows the cumulative probability for all three methods after $n$ images, \textit{given that end of observations is determined only by patch setting and sunrise}. \opin\ and \sein\ cover $51.9\%$ and $50.9\%$ probability by imaging total 103 tiles respectively. \grin\ covers $45.0\%$ probability by imaging 102 tiles, which is $\sim 6.9\%$ less than the \opin\ method. For comparison, \opar, \sear\ and \grar\ cover $53.4\%$, $53.4\%$ $46.9\%$ probability respectively for this patch. Panel (d) shows the probability covered by each of the three algorithms, if observations are to be terminated after $n$ tiles. Each point for \sein\ and \opin\ in curves requires a new run of the algorithm, with the observation end time $t_f$ set at $t_o + n\texp$. We see that \sein\ and \opin\ always outperform the \grin\ algorithm.
    \label{fig:patch_ar}}
\end{figure*}

\subsection{Independent methods}

We now compare the performance of the three ``Independent'' methods: Greedy Independent (\grin), Setting Independent (\sein) and
Optimized Independent (\opin). In general, independent methods perform well when only a small number of observations are possible. Following the discussion in \S\ref{subsec:independent}, we can see that the selection procedures of the three independent methods differ significantly from each other: so the length of the final observing sequence and the location of the selected tiles can vary a lot between these methods.

Since these methods do not use a single grid on the sky, each iteration of \sein\ and \opin\ requires a time-consuming convolution of the tile with the patch. As a result, the computation time for each tile increases proportionally to the size of the localization patch. For the largest patch we simulated (8874~\sqd), the computation of {\em each tile} in the \opin\ observing sequence took about 3~minutes---the same time as a full \opar\ calculation. While these methods are computationally much slower, they are more easy to adapt for usage by an ensemble of telescopes with heterogeneous fields of view. All these calculations can also be sped up significantly by using Fourier--based convolution methods \citep[cf.][]{sps12}, a task we will take up in future work.

We demonstrate the relative performance of the three independent methods using a 313~\sqd\ arc--shaped patch (Figure~\ref{fig:patch_ar}a). On simulating observations with IGO, we find that \opin\ and \sein\ can cover 52\% and 51\% probability respectively. 
The results show qualitative similarities to array methods. In particular, \opin\ starts close to \grin\ in cumulative coverage, drops off as moderate probability regions are about to set, and in the end surpasses \grin, which covers just 45\% probability of containing the true counterpart (Figure~\ref{fig:patch_ar}c). For comparison, \opar\ provides over 53\% probability coverage for this patch. If scheduling constraints warrant that observations terminate earlier, then the \sein\ and \opin\ schedules can be recalculated accounting for this constraint. Figure~\ref{fig:patch_ar}d shows the final result of the three independent methods, assuming observations end after imaging $n$ tiles. We see that in all cases, \opin\ and \sein\ outperform \grin. Further, we find that in all our 1054 simulated patches, \opin\ and \sein\ give similar performance, and are always better than \grin. 

\begin{figure*}[!p]
\centering
    \includegraphics[width=\linewidth]{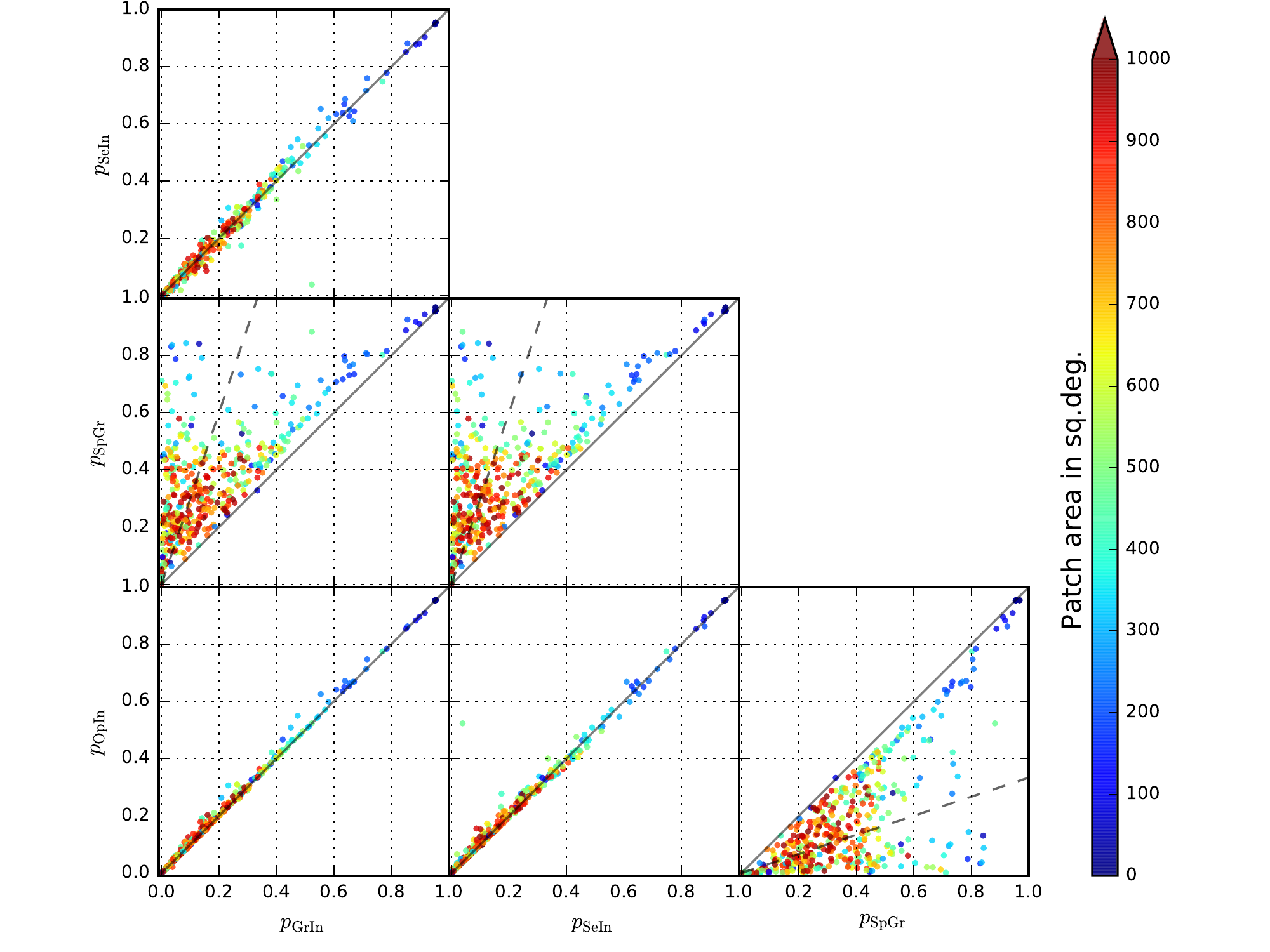}
    \caption{Comparing \opin, \sein, \grin\ and \spgr\ methods. In the online color version, the color of the symbol shows the area of the 95\% patch in square degrees, with red denoting 1000~\sqd\ and larger patches. In each box, the diagonal gray line denotes equal probability coverage by the two methods. On an average, the completely unconstrained \spgr\ algorithm is expected to cover about thrice the probability as ground--based observations. This ratio is denoted by a dashed gray line in the three panels with \spgr\ on one of the axes. We see that \opin\ gives the same or better performance as \grin, while \sein\ may sometimes cover less probability than \grin. \label{fig:all_in}}
\end{figure*}

Lastly, we compare the performance of the three independent methods with each other and \spgr~(Figure~\ref{fig:all_in}). As expected, \opin\ and \sein\ consistently yield better results than \grin, while \spgr\ outperforms all algorithms
in terms of total probability covered. We terminate \spgr\ observations with the last tile of the \opin\
algorithm.

\subsection{Array versus independent}

\begin{figure}[thbp]
\centering
    \includegraphics[trim=0cm 0 1cm 0,clip=true,width=\linewidth]{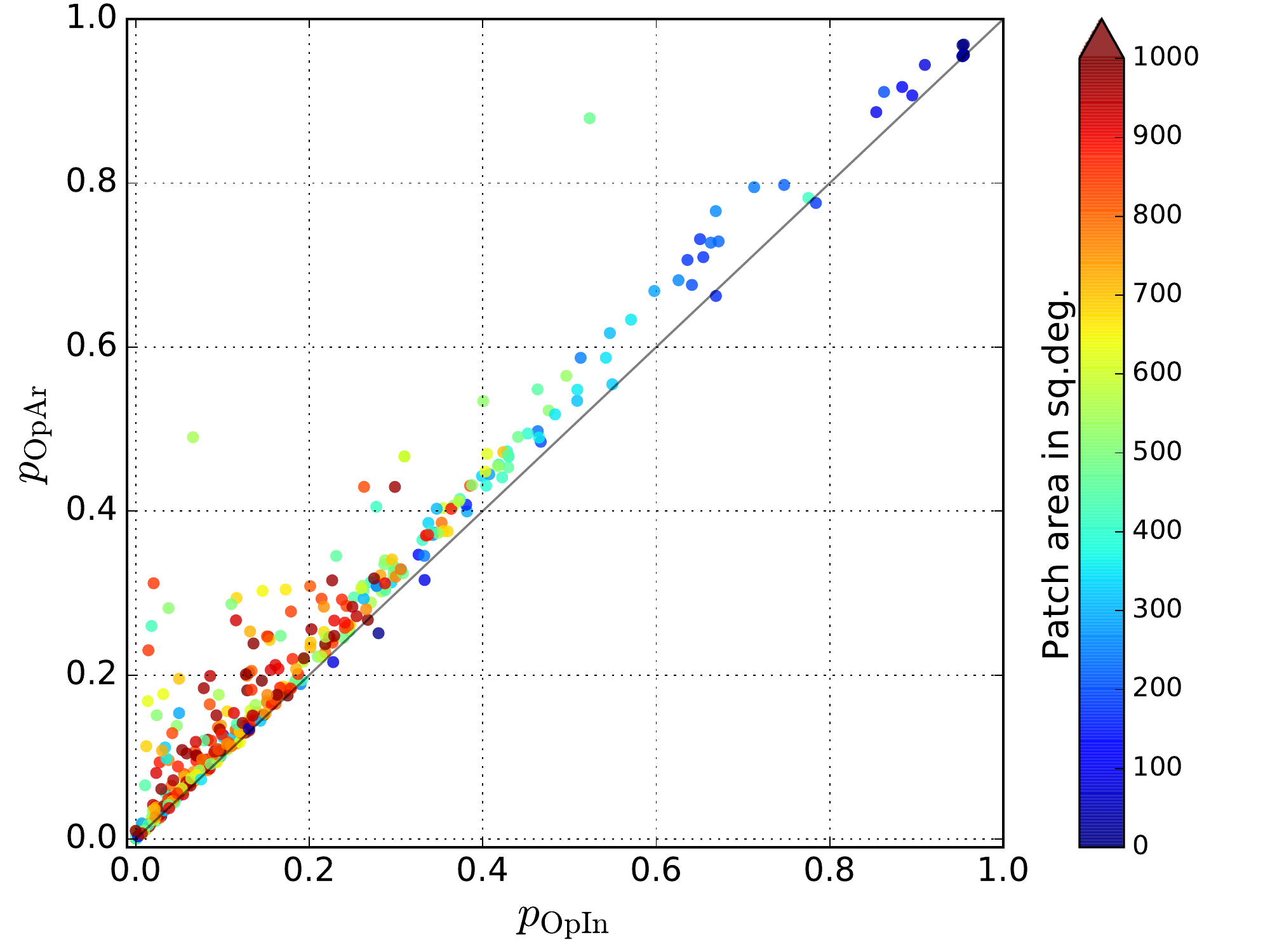}
    \caption{Comparing \opar\ with \opin.  The X--axis and Y--axis are the total probability covered by \opin\ ($p_{\rm OpIn}$) and  \opar\ ($p_{\rm OpAr}$) respectively. In the online color version, the color of the symbol shows the area of the 95\% patch in square degrees, with red denoting 1000~\sqd\ and larger patches. \label{fig:Ar_In_compare}}
\end{figure}
\begin{figure}[htbp]
\centering
    \includegraphics[width=\linewidth]{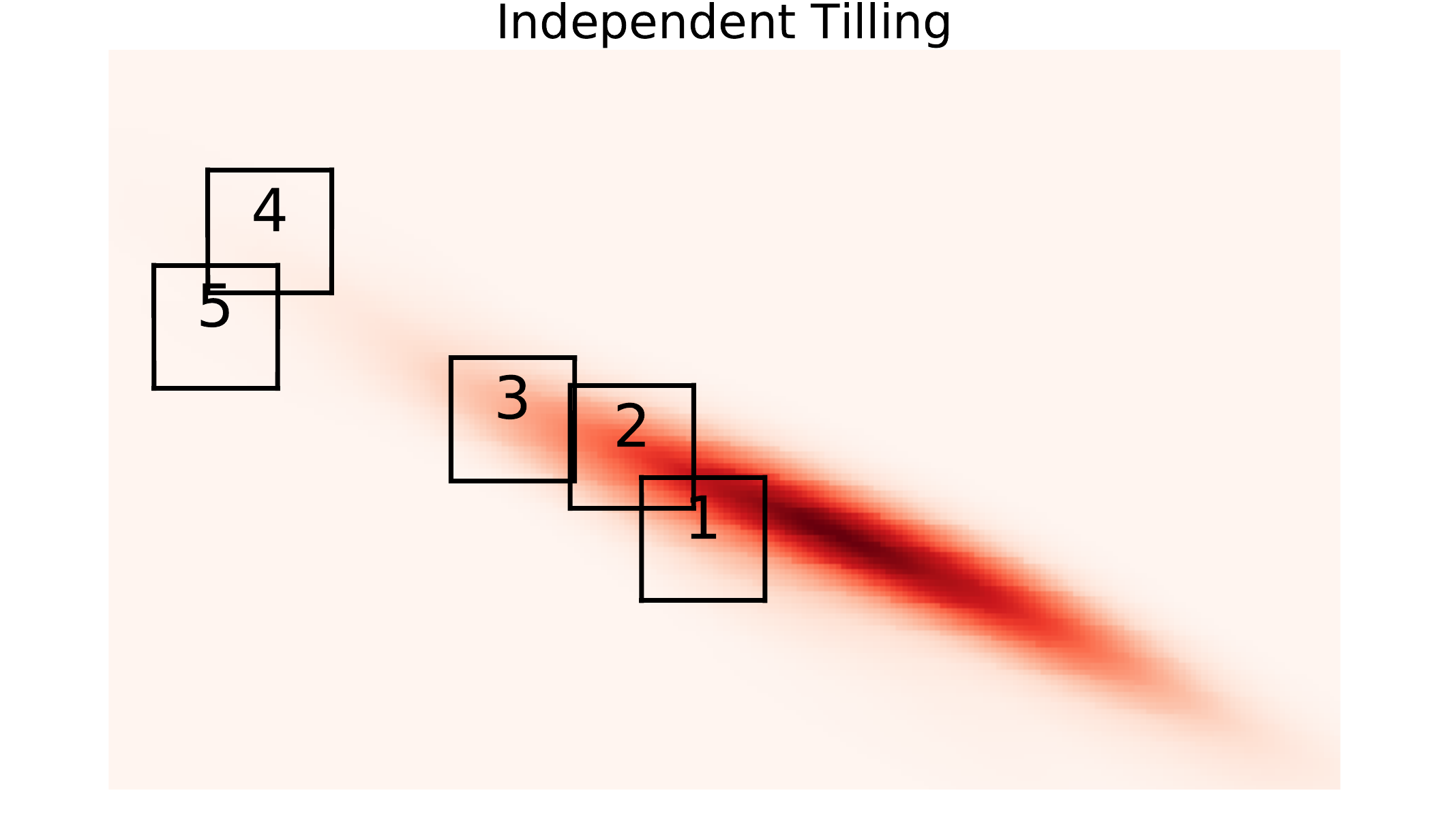}
    \includegraphics[width=\linewidth]{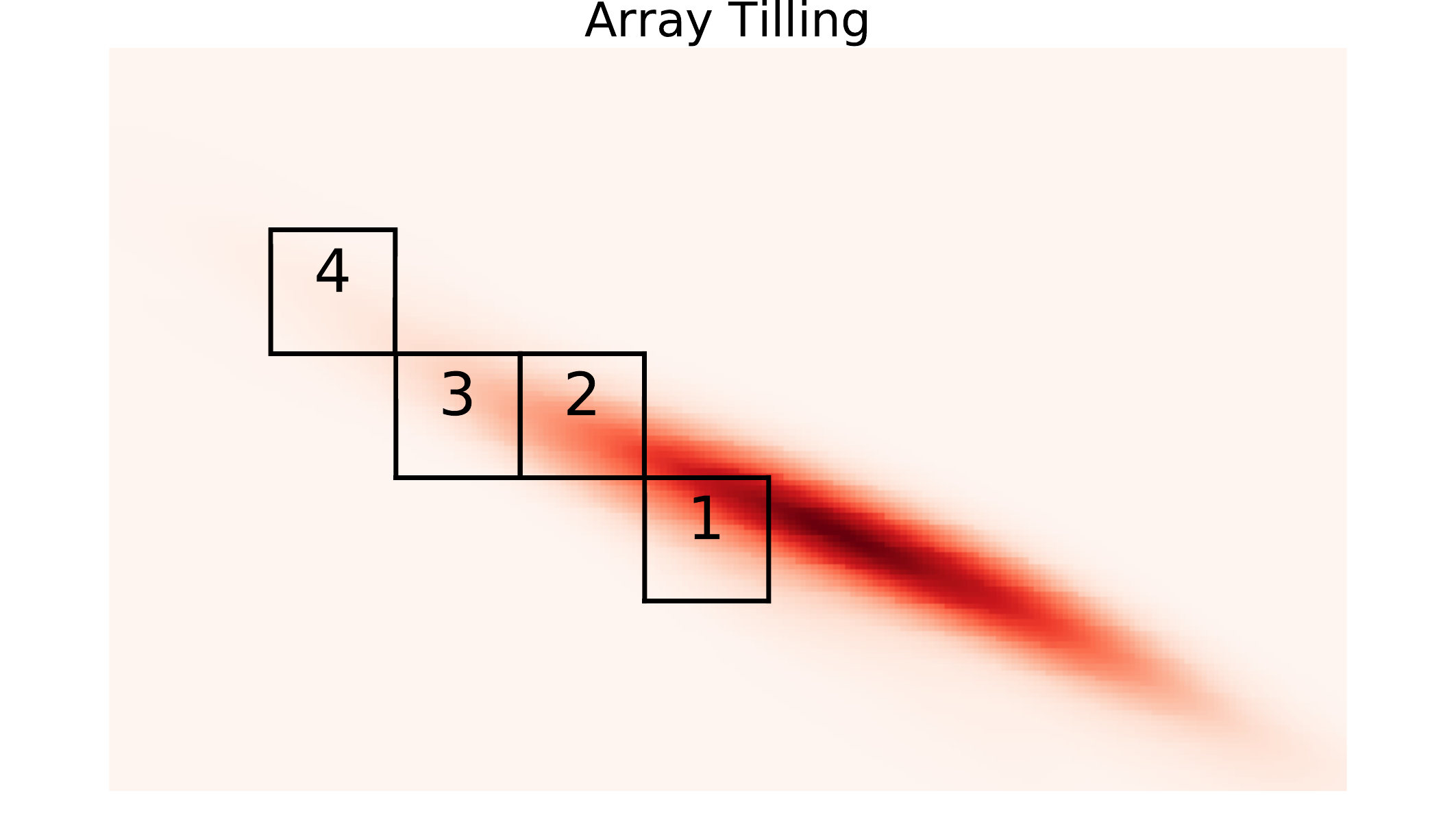}    
    \caption{Tiling plots for a patch of area 14~\sqd, one of the few cases where \opin\ performs better than \opar. It is a setting patch
    and only a few observations are possible. The top plot is for \opin\ and bottom plot is for \opar. In this patch \opin\ covers $28.0\%$ probability by imaging total 5 tiles. \opar\
    covers $25.1\%$ probability by imaging 4 tiles, which is $\sim 2.9\%$ less than the \opin\ method. 
    \label{fig:AlPlt}}
\end{figure}

Comparing our best array and independent methods \opar\ and \opin, we see that array methods usually give higher total probability coverage than independent methods (Figure~\ref{fig:Ar_In_compare}). This stems from irregular gaps left between early \opin\ tiles, which are filled very inefficiently with overlapping observations later. As mentioned in \S\ref{subsec:independent}, independent placement of tiles is primarily useful when only a few observations are possible. One such case is illustrated in Figure~\ref{fig:AlPlt}. This is a relatively small  patch (28~\sqd), which is already setting as observations start. The horizon is to the right in this figure. The first observed tile of both methods is the same. The creation of an array by this first observation forces a sub-optimal selection of subsequent tiles for \opar. \opin, unconstrained by this array, covers higher probability than \opar\ in each successive observation. Finally, the \opin\ has a 28\% probability of having imaged the EM counterpart, in contrast to 25\% for \opar.

\subsection{Field of view trends}

\begin{figure}[htbp]
\centering
    \includegraphics[trim=0cm 0 1cm 0,clip=true,width=\linewidth]{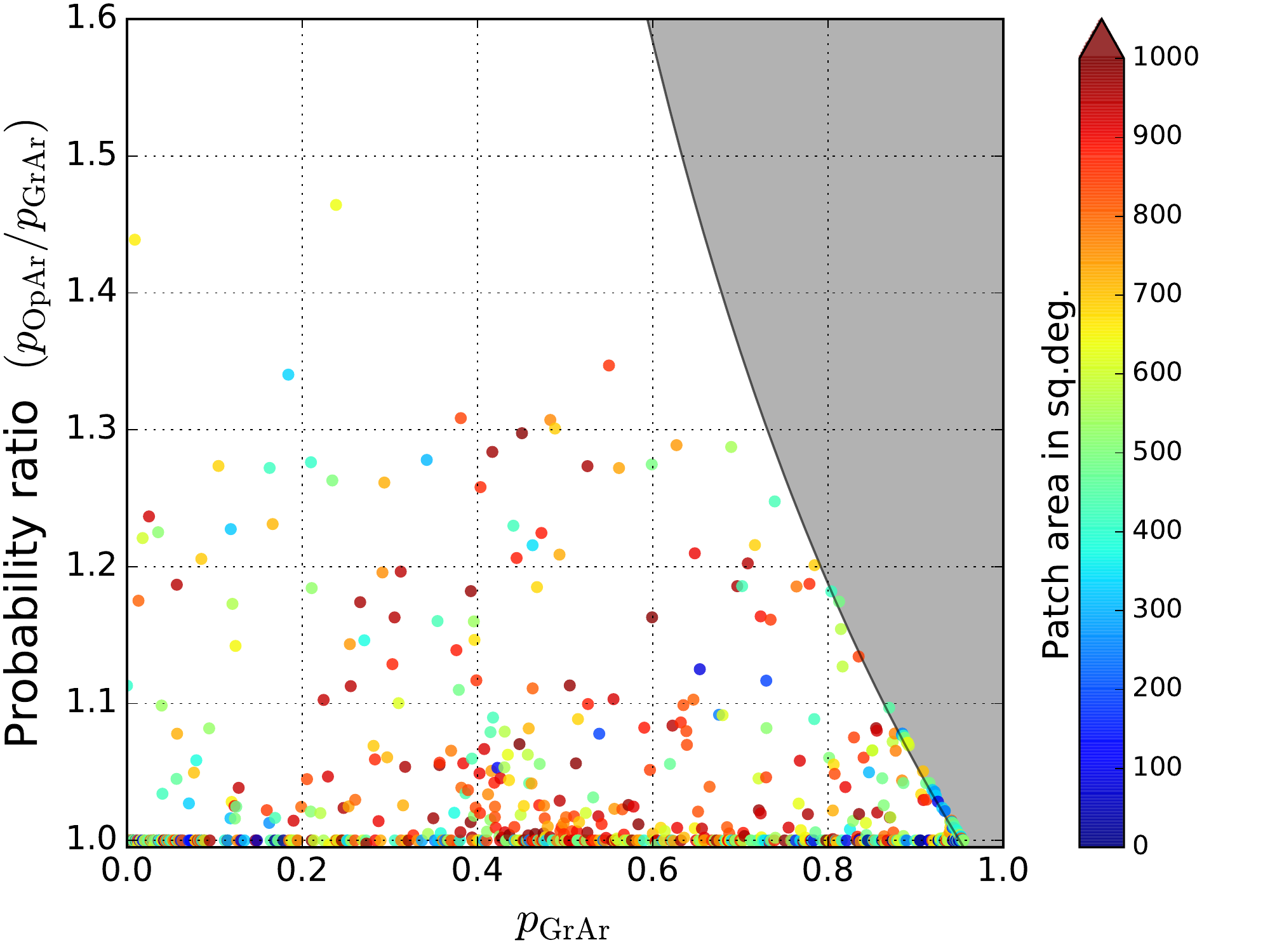}
    \caption{Comparing \opar\ with \grar\ for a telescope with 9~\sqd\ field of view. The X--axis is the total probability covered by \grar\ ($p_{\rm GrAr}$). The Y--axis shows the ratio of total probability covered by \opar\ ($p_{\rm OpAr}$) to $p_{\rm GrAr}$. The shaded gray region on the right is excluded as $p_{\rm OpAr} > 0.95$ or $p_{\rm GrAr} > 0.95$. In the online color version, the color of the symbol shows the area of the 95\% patch in square degrees, with red denoting 1000~\sqd\ and larger patches. \opar\ drastically boosts the probability coverage for many medium and large patches. (Compare the above figure with Fig.~\ref{fig:results_ar}.)
\label{fig:Compare_LFOV}}
\end{figure}

So far, we have simulated observations with a telescope with a $1\degr\times1\degr$ field of view, much smaller than the patch area. In this section, we explore how our results change by using a telescope with a wider field of view. We ran simulations for \opar\ and \grar\ on 1200 patches with areas ranging from few tens of square degrees to over a thousand square degrees, assuming a $3\degr\times3\degr$ square field of view. About two-thirds of these patches (766) were visible from IGO, and we see strong improvements in coverage for many of these.
Figure~\ref{fig:Compare_LFOV} shows the ratio of total probability covered by \opar\ ($p_{\rm OpAr}$) to the total probability covered by \grar\ ($p_{\rm GrAr}$), plotted against $p_{\rm GrAr}$. 
As in Figure~\ref{fig:results_ar} for the smaller field of view, we see that the improvement is small for small patches, but prominent for medium and large patches.

\subsection{Airmass}\label{subsec:airmass}
%

An important consideration in imaging follow--up is the airmass\footnote{Airmass is the relative measurement of the column of air traversed by light from celestial object as it passes through earth's atmosphere. Airmass is unity at the zenith, and increases towards horizon.} at which the tiles are observed: higher airmass leads to lower sensitivity of the telescope. To maximize the chances of finding a counterpart, it is preferable to observe the important high--probability tiles at low airmass, delegating observations of low--probability tiles to high airmass if required. With this in mind, we compare the performance of various methods by using the probability-weighted mean airmass, which we define as follows:
\begin{equation}
\bar{X} = \frac{\sum X_i p_i}{\sum p_i} \label{eq:airmass}
\end{equation}
where the $i^{\rm th}$ tile has probability $p_i$ of containing the EM counterpart, and is observed at airmass $X_i$. (There are, of course, alternative ways of quantifying the airmass but for our purpose it suffices to use Eq.~\ref{eq:airmass}.)

\begin{figure}[htbp]
\centering
    \includegraphics[trim=1cm 0 0cm 0,clip=true,width=\linewidth]{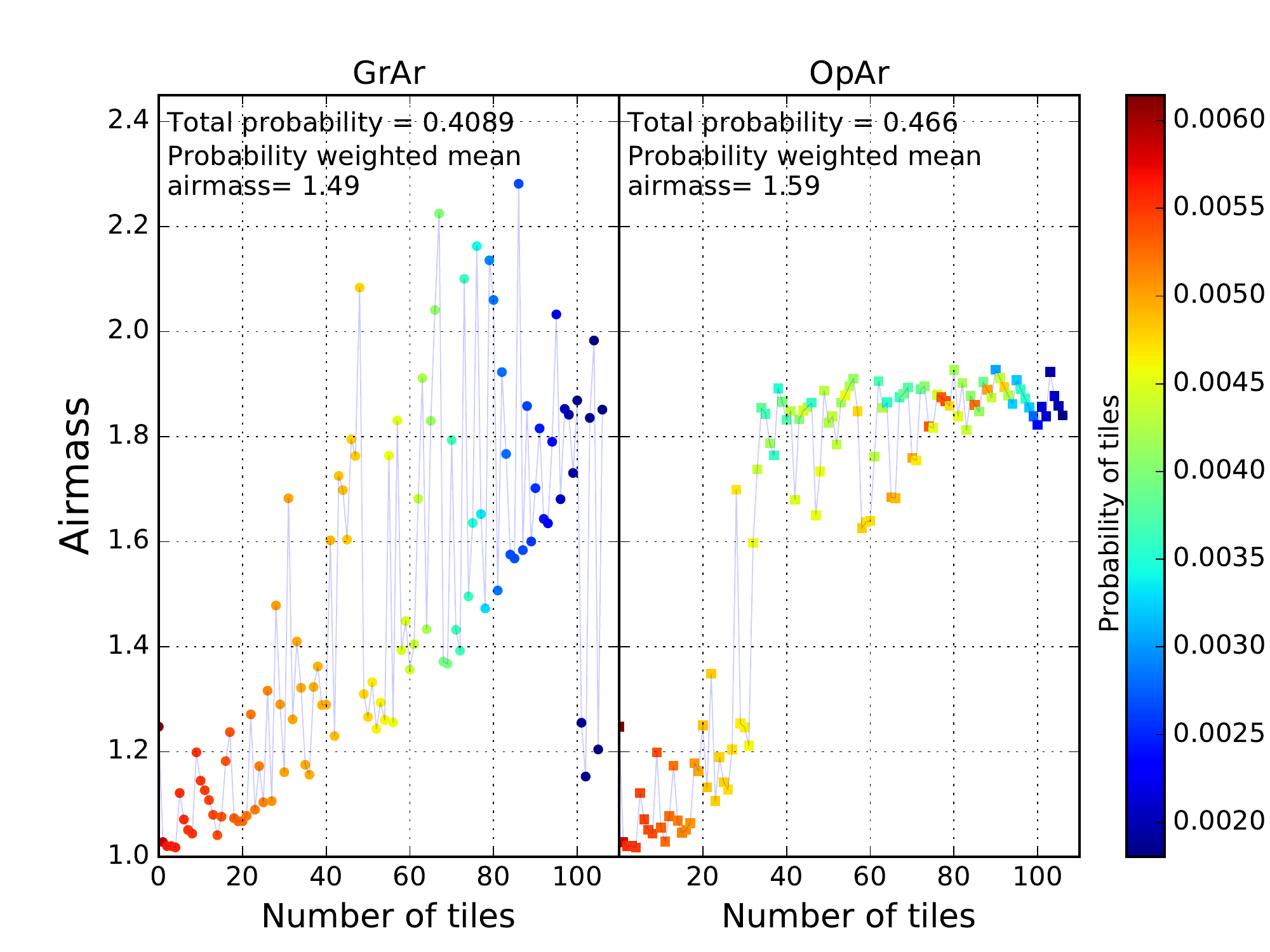}
    \caption{Comparing the airmass of observed tiles in \grar\ and \opar\ schedules for a particular patch. The X--axis is the tile number, and the Y--axis is the airmass at which it is observed. In the online color version, the color of the symbol shows the probability of containing the EM counterpart in each tile. For this patch, the \grar\ schedule shows a lot of jumps in airmass between consecutive tiles, which is not the case with the \opar\ schedule. The probability weighted mean airmass (Equation~\ref{eq:airmass}) for the \grar\ and \opar\ schedules are comparable: 1.5 and 1.6 respectively.\label{fig:Airmass_scatter}}
\end{figure}

\begin{figure}[htbp]
\centering
    \includegraphics[trim=0cm 0 1cm 0,clip=true,width=\linewidth]{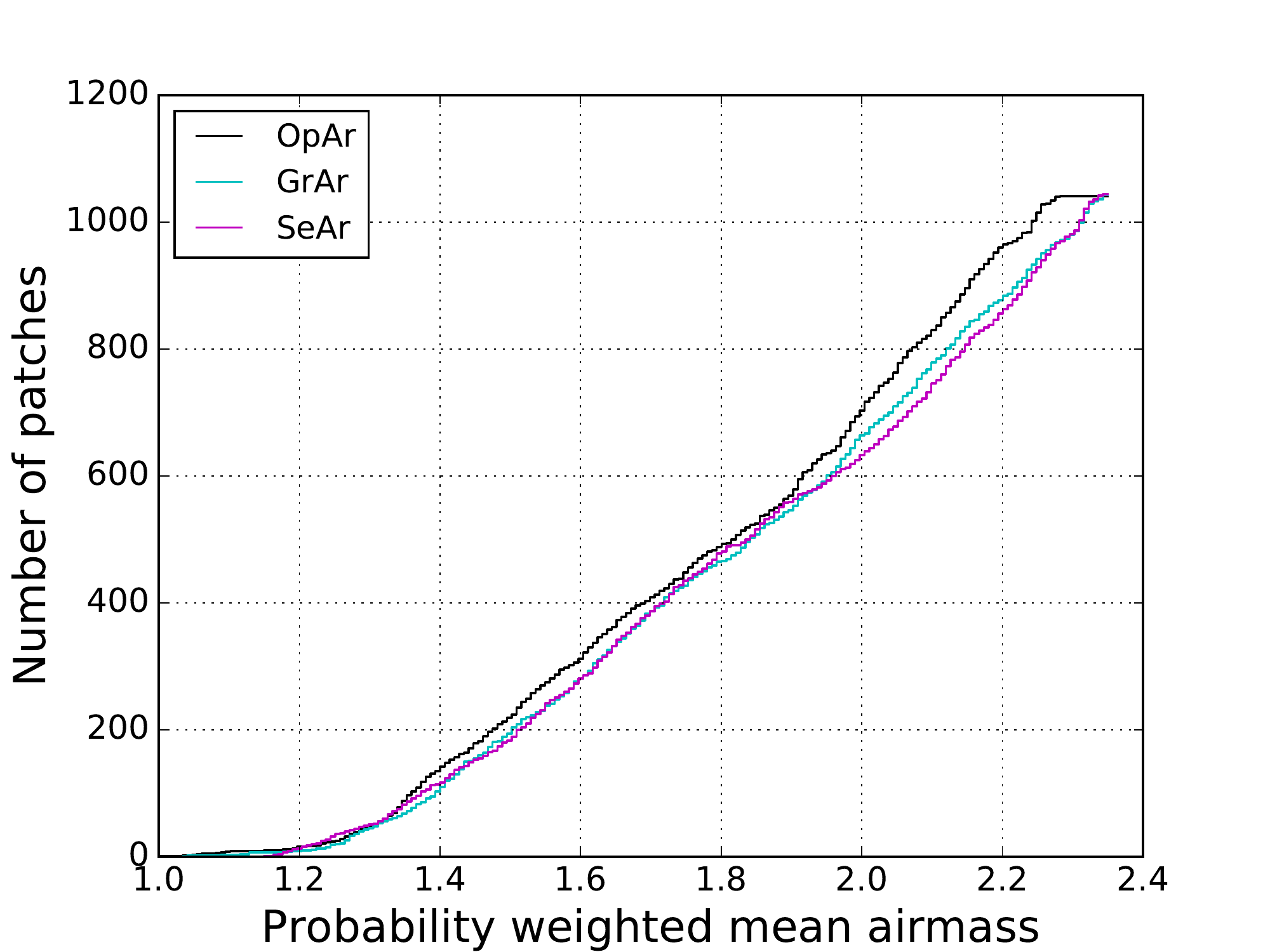}
    \caption{Cumulative distribution of weighted mean airmass ($\bar{X}$, Equation~\ref{eq:airmass}) of observations of 1054 patches with \grar, \sear, and \opar\ schedules. For each airmass on the X--axis, the Y--axis shows the number of patches observed at that or a lower airmass. The three algorithms have comparable airmass performance in our simulations, with \opar\ faring slightly better than the basic \grar.\label{fig:airmass_cumu}}
\end{figure}

One may naively expect that by design our algorithms try to observe tiles just before they set, and may schedule a lot of observations at high airmass. Here, we compare the performance of \opar\ and \grar\ in our simulations. Consider a generic case of observations of a 595~\sqd\ patch scheduled by \opar\ and \grar\ algorithms (Figure~\ref{fig:Airmass_scatter}). The patch happens to be located such that observations end at sunrise. The \grar\ algorithm simply proposes observations of the highest probability tiles in order, following the patch as it sets. As expected, therefore, the typical airmass at which a tile is observed increases with time. The \grar\ schedule for this patch shows a lot of jumps in airmass between consecutive tiles as it jumps to tiles on opposite sides of the patch center, which is not the case with the \opar\ schedule. The default \grar\ algorithm gives 41\% coverage of the probability distribution, and Equation~\ref{eq:airmass} gives a probability-weighted mean airmass of 1.5. By slightly sacrificing the probability-weighted mean airmass ($\bar{X} = 1.6$), \opar\ increases the probability of imaging the EM counterpart to 47\%. However, in general we find that \opar\ schedules have a slightly \textit{better} probability-weighted mean airmass than \grar\ schedules. This is illustrated in the cumulative airmass distributions for three ground--based array algorithms (Figure~\ref{fig:airmass_cumu}). We thus dismiss airmass performance from being an important factor for selecting between these algorithms.

\section{Discussion}\label{sec:discussion}

\subsection{Summary}
In this work, we developed and studied multiple algorithms for calculating observing sequences to image large sky error patches, especially, those of the LIGO-Virgo GW network, by single telescopes or single field of view (FOV) scans. Taking into account day--night and horizon constraints, and restricting observations to a single night, we demonstrate that our \opar\ algorithm always yields the highest probability of imaging the EM counterpart of the transient source. We also showed that there is little penalty in terms of higher airmass observations by selecting the \opar\ algorithm over a default greedy schedule. 

While we considered follow-ups with an optical telescope in this work, it is easily extended to observations with radio telescopes as well. These algorithms can also be adapted for space--based observatories, accounting for solar--, lunar-- and earth--angle constraints. Our simulations end follow--up imaging at the first sunrise after the transient trigger. In case observers want to image more parts of the patch the next night, a new schedule can be computed easily by setting the PDF to zero for all parts of the patch imaged on the first night.

We found that the benefits of using \opar\ increased for larger sky patches. For the 1~\sqd\ FOV used in most of our simulations, \opar\ gave the most enhancement in probability coverage primarily in cases where \grar\ covered $\lesssim 0.3$ probability of imaging the EM counterpart (Figure~\ref{fig:results_ar}). In contrast, for a 9~\sqd\ FOV, \opar\ boosted the chances of imaging the EM counterpart over a wider range of \grar\ probabilities (Figure~\ref{fig:Compare_LFOV}).

We demonstrated that for a majority of cases, observing tiles from a well--defined array gives better coverage than independent placement of all images (Figure~\ref{fig:Ar_In_compare}). In case tiles are to be placed independently, our \opin\ algorithm consistently gives better probability of imaging the EM counterpart as compared to the default \grin\ scheduling algorithm. We found that it is worth calculating an \opin\ observing sequence and comparing it with \opar\ in two cases: (i) when the patch size is relatively small, under about twenty times the tile size; or (ii) only a few images (under a dozen) are possible before the patch sets.

The computation of \opar\ and \sear\ schedules is fast enough for using them in rapid follow--up efforts. Even for patch spanning thousands of square degrees, the schedule is calculated in at most a few minutes on a typical desktop computer (64--bit, 3.6~GHz, 8~GB RAM). On the other hand, implementation of the \opin\ and \sein\ algorithms is computationally intense, as the selection of each tile requires a new convolution of the tile with the patch. For a 1~\sqd\ tile and patches spanning hundreds of square degrees, the calculation can take tens of minutes. We will address this concern in future work.

\subsection{Caveats}
There are some caveats to our current work. We have not considered any lunar constraints in our simulations. These can be incorporated to first order by discarding a circular part of the patch around the moon. However, an accurate implementation should also account for the movement of the moon in the course of the night, which was beyond the scope of this work. Another subtlety is that once a tile sets in our simulations, we do not check if it rises again before sunrise. This could potentially lead to missed opportunities for near--circumpolar tiles, which can set and rise within the observing period. However, considering that such tiles will likely remain at unobservably low altitudes, we expect that ignoring them has negligible effects on our results.

Many synoptic surveys use pre--defined tiles on the sky to ease comparison of old and new images in the search for transients. In our work, the array is placed with one tile centered on the maximum probability density on the sky. Our simulated array can thus be offset from the pre--defined grids of the surveys. As the fields of view used in our simulations are much smaller than the patch sizes, we expect that the small offset arising by replacing our array with the pre--defined grids will not change our results about the efficacy of these algorithms. 

An obvious requirement for counterpart searches is that the telescope must be sensitive enough to detect the counterpart. Throughout this work, we have assumed that a single observation is sensitive enough to detect any EM counterpart that may be present within the field of view imaged by the telescope. If the EM counterpart fades on timescales shorter than a day, it is possible that it may be detectable by a telescope only for the first few hours of observation. Our algorithms can easily incorporate this constraint, by setting the ``observation finish time'' ($t_f$) as the time when the counterpart is expected to be too faint to be detectable. As any such time constraints are model--dependent, we did not incorporate them in this work.


\subsection{Future work}

We are working on applying this algorithm to optimize scheduling when iPTF follows up LIGO triggers~\citep{kcs+16}.

In the near future, we aim to expand our work in two directions. First, transient search programs often prefer obtaining multiple images of each field in a single night. This makes it easy to detect uncatalogued asteroids, which move measurably in tens of minutes. It also allows observers to check for fading of any newly discovered sources. The algorithms proposed here schedule only a single observation per night, and we are evaluating some methods of incorporating multiple epochs each night.

Secondly, we have considered follow--up imaging by a single telescope. For large programs like follow--up of GW sources, it is conceivable that multiple telescopes can work together to increase the overall chance of finding the EM counterpart \citep[cf.][]{sps12}. This adds several complexities to scheduling: (i) each telescope may be able to see a different part of the GW patch, (ii) the observatories may be spread over longitude, hence will start and stop observing at different times, (iii) the fields of view of each telescope may be different, forcing gaps or overlaps in sky coverage, and (iv) the exposure times of each telescope may be different. This creates interesting challenges where the telescopes together have to prioritize between probability--critical tiles (those with high $p_i$ values) and setting--critical tiles (those that will set early). We expect that the \opin\ algorithm, which places each tile independent of previous observations, may be adaptable to such a scenario. These possibilities will be explored in a future work.

\section*{Acknowledgements}
We would like to thank Shaon Ghosh and Mansi Kasliwal for helpful discussions.
One of us (SB) would also like to thank Paul Groot and Gijs Nelemans for their hospitality at Radboud University.  
This work made use of various python libraries and Astropy, the community--developed core Python package for Astronomy. 
This work was supported in part by NSF grants PHY-1206108 and PHY-1506497. 
This document has been assigned the preprint number LIGO-DCC-P1600007.
\software{Numpy~\citep{numpy} and Matplotlib~\citep{matplotlib}, Astropy \citep[\url{http://www.astropy.org}]{astropy}}.

\bibliographystyle{aasjournal}
\bibliography{reference.bib} 

\end{document}